\documentclass[fleqn,usenatbib]{mnras}
\usepackage{newtxtext,newtxmath}
\usepackage{siunitx}
\usepackage{cite}

\usepackage[T1]{fontenc}

\DeclareRobustCommand{\VAN}[3]{#2}
\let\VANthebibliography\thebibliography
\def\thebibliography{\DeclareRobustCommand{\VAN}[3]{##3}\VANthebibliography}

\usepackage{graphicx}	
\usepackage{amsmath}	

\usepackage{subfig}         
\usepackage{overpic}

\newcommand{\msun}{M$_{\odot}$}

\newcommand{\oiii}{\hbox{[O\,{\sc iii}]}}
\newcommand{\ha}{\hbox{H$\alpha$}}
\newcommand{\hb}{\hbox{H$\beta$}}
\newcommand{\oii}{\hbox{[O\,{\sc ii}]}}
\newcommand{\oi}{\hbox{[O\,{\sc i}]}}
\newcommand{\sii}{\hbox{[S\,{\sc ii}]}}
\newcommand{\nii}{\hbox{[N\,{\sc ii}]}}
\newcommand{\heii}{\hbox{He\,{\sc ii}}}

\newcommand{\eddratio}{$\lambda_{\rm{Edd}}$}

\newcommand{\loiii}{$L_{\rm{[OIII]}}$}
\newcommand{\sigmasfr}{${\Sigma_{\rm SFR}}$}
\newcommand{\sigmasfrkpc}{${\Sigma_{\rm SFR,1 kpc}}$}

\newcommand{\sigmaoiiikpc}{${\Sigma_{\rm {[OIII]},1 kpc}}$}

\newcommand{\sfrkpc}{$\rm SFR_{1 kpc}$}
\newcommand{\ssfrkpc}{$\rm sSFR_{1 kpc}$}
\newcommand{\sigmaha}{$\rm \Sigma_{H\alpha}$}

\usepackage{hyperref}
\usepackage{academicons}
\usepackage[dvipsnames]{xcolor}
\usepackage{orcidlink}

\title[Galaxies with Biconical Ionized Structure]{Galaxies with Biconical Ionized Structure in MaNGA - I. Sample Selection and Driven Mechanisms}

\author[Zhijie Zhou et al.]{
\parbox[t]{\textwidth}{\raggedright
Zhi-Jie Zhou$^{1,2,3}$~\orcidlink{0000-0003-1709-6005}, 
Yan-Mei Chen$^{1,2,3}$\thanks{E-mail: \href{mailto:chenym@nju.edu.cn}{chenym@nju.edu.cn}}~\orcidlink{0000-0003-3226-031X}, 
Run-Quan Guan$^{4}$~\orcidlink{0000-0002-9868-0838}, 
Yong Shi$^{1,2,3}$~\orcidlink{0000-0002-8614-6275}, 
Qiu-Sheng Gu$^{1,2,3}$~\orcidlink{0000-0002-3890-3729},
and 
Dmitry Bizyaev$^{5,6}$~\orcidlink{0000-0002-3601-133X}
}\\
\vspace*{6pt}\\
$^{1}$School of Astronomy and Space Science, Nanjing University, Nanjing 210093, China\\
$^{2}$Key Laboratory of Modern Astronomy and Astrophysics (Nanjing University), Ministry of Education, Nanjing 210093, China\\
$^{3}$Collaborative Innovation Center of Modern Astronomy and Space Exploration, Nanjing 210093, China\\
$^{4}$University of California at Santa Cruz, Santa Cruz, CA 95064, USA\\
$^{5}$Apache Point Observatory and New Mexico State University, P.O. Box 59, Sunspot, NM 88349-0059, USA\\
$^{6}$Sternberg Astronomical Institute, Moscow State University, Moscow, Russia\\
}

\date{Accepted 2024 May 03. Received 2024 February 26; in original form 2023 September 28 }

\pubyear{2024}

\begin{document}
\label{firstpage}
\pagerange{\pageref{firstpage}--\pageref{lastpage}}

\maketitle

\begin{abstract} \label{abstract}
	Based on the integral field unit (IFU) data from Mapping Nearby Galaxies at Apache Point Observatory (MaNGA) survey, we develop a new method to select galaxies with biconical ionized structures, 
	building a sample of 142 edge-on biconical ionized galaxies. We classify these 142 galaxies into 81 star-forming galaxies, 31 composite galaxies, and 30 AGNs (consisting of 23 Seyferts and 7 LI(N)ERs) according to the {\nii}-BPT diagram.  
	The star-forming bicones have bar-like structures while AGN bicones display hourglass structures, 
	and composite bicones exhibit transitional morphologies between them due to both black hole and star-formation activities. 
	Star-forming bicones have intense star-formation activities in their central regions, and the primary driver of biconical structures is the central star formation rate surface density. 
	The lack of difference in the strength of central black hole activities (traced by dust attenuation corrected {\oiii}$\lambda$5007 luminosity and Eddington ratio) between Seyfert bicones and their control samples 
	can be naturally explained as that the accretion disk and the galactic disk are not necessarily coplanar. 
	Additionally, the biconical galaxies with central LI(N)ER-like line ratios are edge-on disk galaxies that show strong central dust attenuation. 
	The radial gradients of {\ha} surface brightness follow the $r^{-2.35}$ relation, roughly consistent with $r^{-2}$ profile, which is expected in the case of photoionization by a central point-like source. 
	These observations indicate obscured AGNs or AGN echoes as the primary drivers of biconical structures in LI(N)ERs. 

\end{abstract}


\section{Introduction}

Feedback plays a critical role in regulating the formation and evolution of galaxies \citep{1997ApJ...481..703S}, shaping the properties of galaxies \citep[such as star formation history, morphology;][]{2009ApJ...707..250M}. 
Feedback also influences the distribution and properties of the interstellar medium (ISM) as well as circumgalactic medium (CGM) \citep{1990ApJS...74..833H,2017ARA&A..55..389T}. 
In order to realistically reproduce observational results (such as the galaxy luminosity function), numerical simulations incorporate feedback process into galaxy evolution models to suppress star formation \citep{2006MNRAS.370..645B,2006MNRAS.365...11C}. 
For the low luminosity end, supernova (SNe) feedback is applied, while active galactic nucleus (AGNs) feedback is employed for the high luminosity end \citep{2003MNRAS.339..289S,2005MNRAS.361..776S}.

Galactic-scale outflow is an extremely spectacular form of feedback. It is known to be ubiquitous in the actively star-forming galaxies with $\rm \Sigma_{SFR} = 0.1\ M_{\odot}\ yr ^{-1}\ kpc^{-2}$ \citep{1980A&A....87..152H,1990ApJS...74..833H,2009ApJ...692..187W,2013Natur.499..450B,2021ApJ...919....5B} 
as well as some AGNs \citep{2009MNRAS.397..249P,2012MNRAS.425..605F,2012MNRAS.420.2221D,2015ApJ...799...83C}. 
A direct-viewing image of galactic-scale wind is given by the nearby starburst galaxy M82 \citep{1987PASJ...39..685N}, 
where the multi-phase outflow is observed by different space telescopes, including X-ray hot gas from Chandra \citep{2007ApJ...658..258S}, 
ionized gas ({\ha} emission) traced by Hubble Space Telescope \citep{2007PASP..119....1M}, and $8 \mu m$ band images (tracing cool gas and dust) from Spitzer Space Telescope \citep{2006ApJ...642L.127E}. 
The outflow extends over kiloparsecs scale, and its velocity reaches to a few hundred kilometers per second \citep{1990ApJS...74..833H}. 
However, such kind of observations need to mobilize multiple space telescopes, the cost is expensive and can be only applied to several nearby galaxies. 

Recent development in IFU technology has promoted spatially resolved galaxy surveys, such as the Calar Alto Legacy Integral Field Area \citep[CALIFA;][]{2012A&A...538A...8S} Survey, 
the Sydney-AAO Multi-object Integral field spectrograph \citep[SAMI;][]{2012MNRAS.421..872C}, 
and the Mapping Nearby Galaxies at Apache Point Observatory \citep[MaNGA;][]{2015ApJ...798....7B} survey. 
These spatially resolved data can provide not only information about the structure of the outflows, but also their kinematics. 
\citet{2019MNRAS.490.3830B} discovered an edge-on Seyfert 2 galaxy with X-shaped biconical outflow from MaNGA survey. 
The counter-rotating gas and stellar components in this galaxy provide an effective mechanism for the loss of angular momentum, 
leading to the gas flow and the central black hole accretion, which is followed by the AGN-driven galactic-scale wind. 
\citet{2022ApJ...925..203J} presented a case study of the nearby galaxy NGC 7582 using Multi Unit Spectroscopic Explorer (MUSE) observation, 
obtaining a clear view of the outflowing cones over kiloparsec scales as well as gas kinematics inside the cones. 
Thanks to the unprecedented large number of spatial resolved sample given by the MaNGA survey, statistical studies on galaxy samples with biconical gas outflow become possible. 
\citet{2019ApJ...882..145B,2022MNRAS.516.3092B} built samples of nearby star-forming galaxies with biconical outflow, finding these galaxies have high central concentration of the star formation rate, 
which increases the gas velocity dispersion over the equilibrium limit and helps maintain the gas outflows. 
Again based on MaNGA data, \citet{2021MNRAS.505..191B} studied 36 biconical star-forming galaxies finding that they have enhanced star formation rate and metallicity along photometric minor axis than major axis, 
indicating a positive feedback and metal entrainment in the galactic-scale outflows.

In this study, we develop a method to systematically identify edge-on galaxies with biconical ionized structures from the SDSS DR17 data and investigate their primary driving mechanisms. 
In Section 2, we give a short introduction to the MaNGA survey, and the detailed description of sample selection method. 
In Section 3, we separate the sample into three subgroups (star-forming galaxies, composite galaxies, and AGNs) according to {\nii}-BPT diagnosis diagram \citep{1981PASP...93....5B}, 
studying the morphologies and primary driving mechanisms of biconical structures. 
In Section 4, we analyze the possible origin of energy buget which is required in driving biconical structures in LI(N)ER host galaxies. 
Finally, we give a short summary in Section 5. 
In this paper, we adopt the $\Lambda$ cold dark matter model with $H_{0} = 70\rm{km\ s^{-1} Mpc^{-1}}$, $\rm{\Omega_{m}} = 0.30$, and $\rm{\Omega_{\Lambda}} = 0.70$.

\section{Data reduction and sample selection}

\subsection{The MaNGA survey}

The MaNGA survey \citep{2015ApJ...798....7B} is one of the three major programs of the fourth-generation Sloan Digital Sky Survey (SDSS-IV).
It uses the 2.5-m Sloan Foundation Telescope \citep{2006AJ....131.2332G} at Apache Point Observatory (APO) to observe a sample of 10,010 unique galaxies comprised of all morphological types \citep{2022ApJS..259...35A}. 
The targets cover an almost flat distribution in stellar mass ($M_{*}$) with a range of $10^{9}$ {\msun} $< M_{*} < 10^{11}$ {\msun}, and a redshift interval $0.01 < z < 0.15$ \citep{2017AJ....154...28B}. 
MaNGA uses the IFU spectroscopy technique, observing each galaxy with 19-127 hexagonal fiber bundles, corresponding to $12.5\arcsec - 32.5 \arcsec$ diameter in the sky, 
depending on the apparent size of the target. 
MaNGA observes two-thirds of the galaxies with fields of view covering 1.5 times the effective radius ($R_{\rm e}$), which contains half of the galaxy light in $r$-band, 
and the other one-third have coverage out to 2.5$R_{\rm e}$. They are referred as primary and secondary samples respectively. 
The two dual-channel BOSS spectrographs \citep{2013AJ....146...32S} provide simultaneous spectra that have a point-spread function (PSF) with a full width at half-maximum (FWHM) of $\sim2.5 \arcsec$, 
and wavelength coverage ranges from $3,600 - 10,300$ {\AA} with a median spectral resolution $R\sim2,000$. 
More information on the observing strategy, calibration, and survey design of MaNGA can be found in \citet{2015AJ....150...19L}, \citet{2016AJ....151....8Y}, and \citet{2017AJ....154...86W}.

\subsection{Data analysis}
In this work, we apply data from the SDSS DR17, which includes $10,010$ unique galaxies. 
The MaNGA Data Analysis Pipeline \citep[DAP;][]{2019AJ....158..160B} utilized {\tt pPXF} \citep{2004PASP..116..138C} and a combination of the MILES stellar library \citep{2006MNRAS.371..703S} for the kinematics and the MaStar SSP library \citep{2022ApJS..259...35A} for the stellar continuum. 
The stellar continuum fitting process produces estimates of stellar absorption lines, alongside measurements of 21 major nebular emission lines over MaNGA wavelength coverage.  
Eventually these parameters are stored in MaNGA DAP products named `SPX-MILESHC-MASTARSSP', which allow us to obtain the spatially resolved 4000\AA\ break ($\rm D$$_{n}4000$), 
line-of-sight stellar rotation velocity ($V_{*}$), stellar velocity dispersion ($\sigma_{*}$), line-of-sight rotation velocity of ionized gas ($V_{\rm gas}$), 
equivalent width of emission lines, including {\oiii}$\lambda$5007 ($\rm EW_{[OIII]}$), {\ha} ($\rm EW_{\rm H\alpha}$) et al. 

We measure the global $\rm D$$_{n}4000$ from stacked spectra of each galaxy. 
$\rm D$$_{n}4000$ is defined as the flux ratio between two narrow bands of $4000-4100$ \AA \ and $3850-3950$ \AA, which is a good tracer of the light-weighted stellar population age. 
We stack all spectra with median signal-to-noise ratio ($\rm S/N$) per spaxel greater than 2 within the MaNGA bundle to measure the global $\rm D$$_{n}4000$. 
The global stellar masses, effective radius of the galaxies are cross-matched from the NASA Sloan Atlas (NSA) catalog \citep{Blanton_2011}. 
The global stellar masses are estimated from spectral energy distribution (SED) fitting by the {\tt kcorrect} software package \citep{2007AJ....133..734B} with BC03 simple stellar population models \citep{2003MNRAS.344.1000B} and \citet{2003PASP..115..763C} initial mass function. 

The SDSS images are observed through CCD imaging in $u, g, r, i, z-$bands \citep{1996AJ....111.1748F,2002AJ....123.2121S}, using a drift scan camera mounted on a wide-field 2.5-m telescope \citep{1998AJ....116.3040G}. 
The photometric pipeline \citep{2001ASPC..238..269L} fits each galaxy image with a two-dimensional model of a de Vaucouleurs and an exponential surface profile, taking into account the PSF. 
The pipeline determines the optimal linear combination of the exponential and de Vaucouleurs functions and saves it as a parameter named fracDeV \citep{2004AJ....128..502A}. 
$\rm fracDeV = 1$ corresponds to a pure de Vaucouleurs profile, while $\rm fracDeV = 0$ corresponds to an exponential disk.
Following \citet{2008MNRAS.388.1321P}, we use fracDeV to distinguish disk galaxies ($\rm fracDeV < 0.8$) from elliptical galaxies ($\rm fracDeV \geqslant 0.8$). 

We obtained the $r$-band fracDeV, absolute magnitude ($M_{\rm r}$), and axis ratio ($b/a$) from the SDSS PhotoObjAll table in DR17\footnote{\href{https://skyserver.sdss.org/dr17/}{SDSS SkyServer DR17}}. 
Following the method proposed by \citet{2008MNRAS.388.1321P}, we estimate the inclination of 5593 disk galaxies from $b/a$ and $M_{\rm r}$. 
The inclination ($i$) ranges from $\ang{0}\sim\ang{90}$, with $\ang{0}$ representing purely face-on galaxies while $\ang{90}$ representing extremely edge-on galaxies. 
Among the 4417 galaxies with $\rm fracDeV \geqslant 0.8$, we classified them as elliptical galaxies and set their inclinations to $i = -\ang{200}$.

\subsection{Sample Selection}

\begin{figure*}
	\centering
	\includegraphics[width=2\columnwidth]{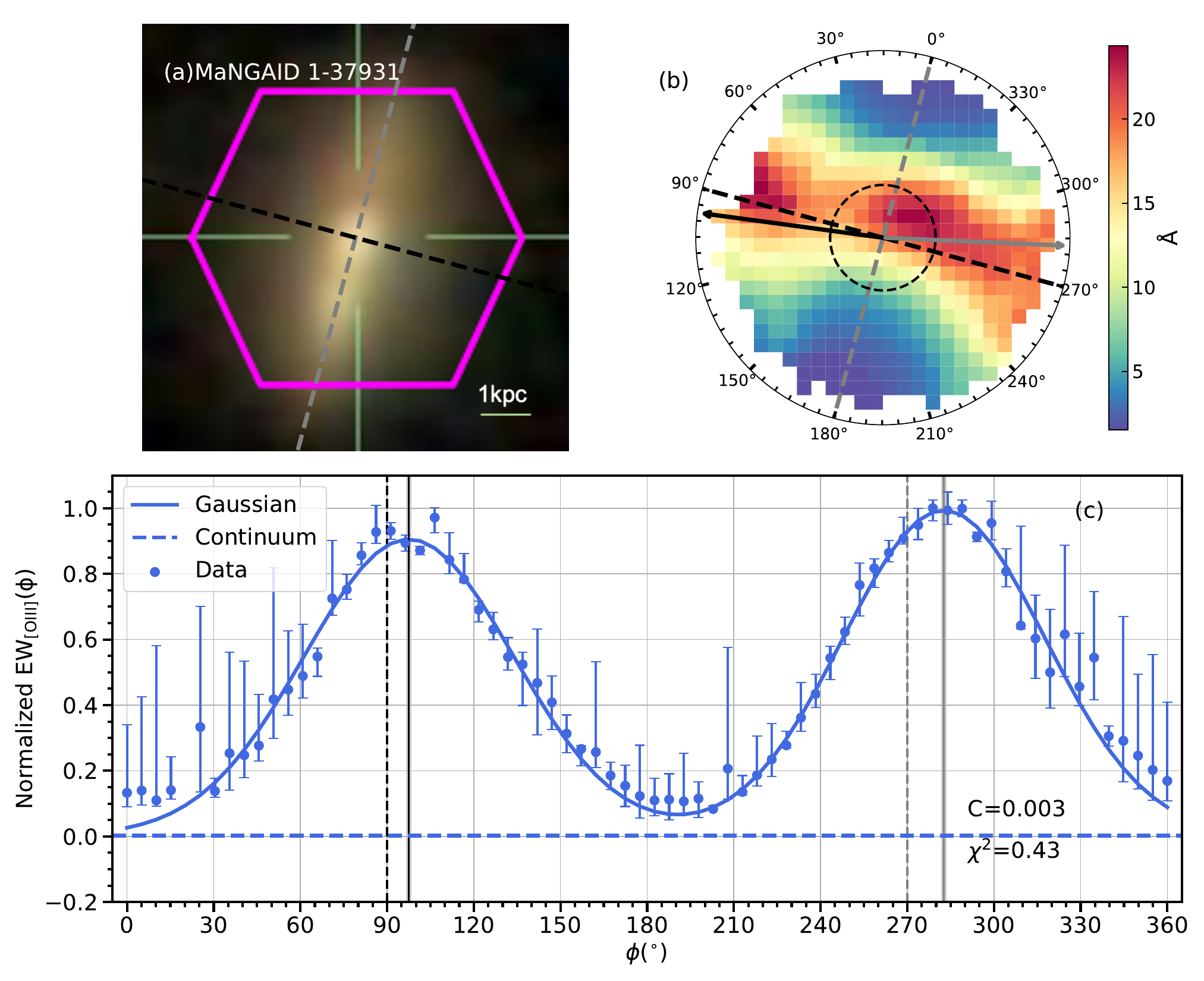}
    \caption{
	{\oiii}$\rm \lambda$5007 equivalent width distribution of a galaxy with the biconical ionized structures. 
	(a) SDSS $g, r, i$ color image of the galaxy. The magenta hexagon marks the MaNGA bundle.
	(b) The {\oiii}$\rm \lambda$5007 equivalent width map, $\phi$ increases counterclockwise from the half major axis. 
		The gray dashed line shows the photometric major axis, while the black dashed line shows the photometric minor axis. 
		The arrows correspond to the double-Gaussian peaks. 
		Black dashed circle represents an aperture of 1 kpc radius. 
	(c) {\oiii}$\rm \lambda$5007 equivalent width as a function of $\phi$. The blue dots are the observations, and the error bars cover $\rm \pm 1\sigma$ range. 
	The blue curve shows the best-fit double-Gaussian model, and the blue dashed line represent the continuum level of the $\rm EW_{[OIII]}$. 
	The black and gray dashed lines show the positions of minor axes ($\ang{90}$ and $\ang{270}$). 
	The black and gray solid lines mark the peak positions of each Gaussian component, which are consistent with the arrows in panel (b), and the grey shadows cover $\rm \pm 1\sigma$ range. 
	}
    \label{fig:angular_distribution}
\end{figure*}

\begin{figure*}
	\centering
	\includegraphics[width=2\columnwidth]{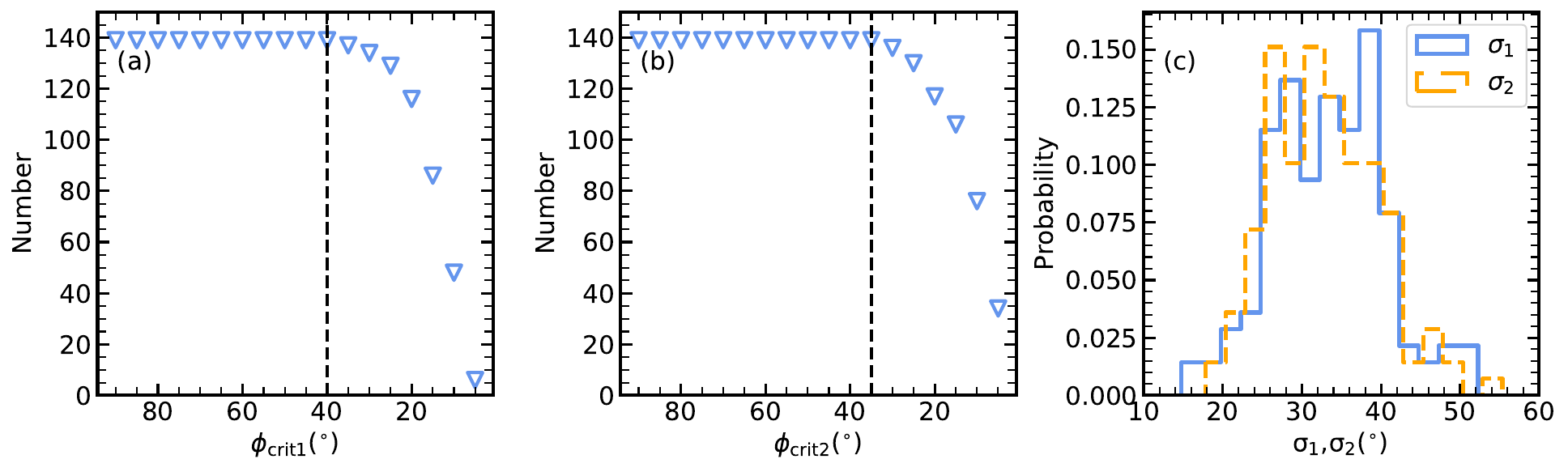}
    \caption{
	(a) The number of biconical galaxies selected by eye as a function of $\phi_{\rm crit1}$ once we apply sample selection criterion (3) to the sample. 
	(b) The number of biconical galaxies selected by eye as a function of $\phi_{\rm crit2}$ once we apply sample selection criterion (4) to the sample. 
	(c) $\rm \sigma_{1}$, $\rm \sigma_{2}$ distributions of biconical galaxies selected by eye. 
	}
    \label{fig:criteria}
\end{figure*}

Galactic-scale winds are believed to be driven by the feedback from either star formation or black hole activities. 
In actively star-forming galaxies, galactic winds are triggered by the mechanical energy and momentum released by SNe and stellar winds \citep{1985Natur.317...44C,1990ApJS...74..833H}. 
Young star clusters create over pressured bubbles of hot gas that expand and sweep up surrounding ISM. 
The collective action of multiple super bubbles leads to a weakly collimated biconical outflow along the minor axis of a galaxy once they ``blow out'' from the disk into the halo \citep{2003RMxAC..17...47H,Chen_2010,2014ApJ...797...90S}. 
The outflow consists of hot gas and cool entrained clouds. 
In host galaxies of AGNs, the formation picture of galactic wind is similar to that of star-forming galaxies, 
the only difference is that the galactic wind is generated by one bubble driven by the activity of the central supermassive black hole \citep{doi:10.1146/annurev-astro-082214-122316,2019MNRAS.490.3025R}. 

In this work, we are interested in the biconical structure of the ionized gas. In the optical band, the ionized gas is primarily traced by the emission lines, such as {\ha}, {\oiii}$\lambda5007$. 
As the first step of sample selection, we separate the emission-line galaxies from those with weak or no emission lines. 
We define emission-line galaxies as those with at least $10\%$ spaxels with $\rm S/N$ greater than 3 for {\oiii}$\lambda5007$ and {\ha} emission lines. 
Based on this criterion, we select 7577 out of 10010 sources as emission-line galaxies, and our subsequent analysis is based on these galaxies. 

In both \citet{2019MNRAS.490.3830B} and \citet{2022ApJ...925..203J}, they analyze IFU data of individual galaxy with biconical ionized structure, 
pointing out that {\ha} emission is dominant in the disk while {\oiii}$\lambda5007$ dominated in the ionization cones (see Fig.5 of \citealt{2019MNRAS.490.3830B}, Fig.7 of \citealt{2022ApJ...925..203J}).  
Considering that the emission lines from the ionization cone is much weaker than that from the galactic disk, it is overshined by the disk component in face-on systems. 
Thus, the idea of our sample selection method is searching for characteristics of biconical ionized structure in $\rm EW_{[OIII]}$ maps along photometric minor axes in edge-on galaxies where the ionized cones are more easily to be observed. 

Fig.~\ref{fig:angular_distribution} presents an example of galaxy with ionization cones. 
Fig.~\ref{fig:angular_distribution}(a) shows the SDSS $g, r, i$-band image. 
Fig.~\ref{fig:angular_distribution}(b) displays the $\rm EW_{[OIII]}$ map, 
the half major axis is set at $\phi = \ang{0}$, with increasing values of $\phi$ in the counter-clockwise direction.
To analyze the $\rm EW_{[OIII]}$ distribution, we divide the map into circular sectors with a sector width of $\Delta \phi = \ang{5}$. 
Fig.~\ref{fig:angular_distribution}(c) shows the $\rm EW_{[OIII]}$ as a function of $\phi$, with the peak value normalized to 1, and the blue dots are the median value of $\rm EW_{[OIII]}$ in each sector. 
We fit two Gaussian components to Fig.~\ref{fig:angular_distribution}(c) to search for the $\rm EW_{[OIII]}$ enhanced regions as
\begin{equation*}
	\rm EW_{\rm [OIII]} (A_{\rm{i}},\phi_{\rm {i}},\sigma_{\rm{i}},C) = A_{1} e ^{- \frac{(\phi-\phi_{1})^2}{2 {\sigma_{1}}^2}} + A_{2} e ^{- \frac{(\phi-\phi_{2})^2}{2 {\sigma_{2}}^2}} + C,
\end{equation*}
where $A_{\rm 1}$,$A_{\rm 2}$ represents the amplitude of each Gaussian component, $\phi_{\rm 1}$,$\phi_{\rm 2}$ denotes the peak position of each Gaussian, $\sigma_{\rm 1}$,$\sigma_{\rm 2}$ corresponds to the width of the Gaussian distribution, 
and $C$ represents the continuum level of $\rm EW_{[OIII]}$, namely the intensity of $\rm EW_{[OIII]}$ outside biconical structures. 
The blue curve in Fig.~\ref{fig:angular_distribution}(c) is our best-fit Gaussian model with a continuum $C = 0.003$ shown as the blue dashed horizon line, 
which is fitted with the {\tt curve\_fit} function in {\tt PYTHON}, employing the non-linear least squares method. 
The best fitting parameters are determined by minimizing the reduced $\chi^{2}$ value. 
The two peaks of $\phi_{1} = 97.9\pm \ang{0.73}$ and $\phi_{2} = 283.4\pm \ang{0.68}$ are marked by black and gray solid vertical lines, which corresponds to the black and gray arrows in Fig.~\ref{fig:angular_distribution}(b).

\begin{figure*}
	\centering
	\includegraphics[width=2\columnwidth]{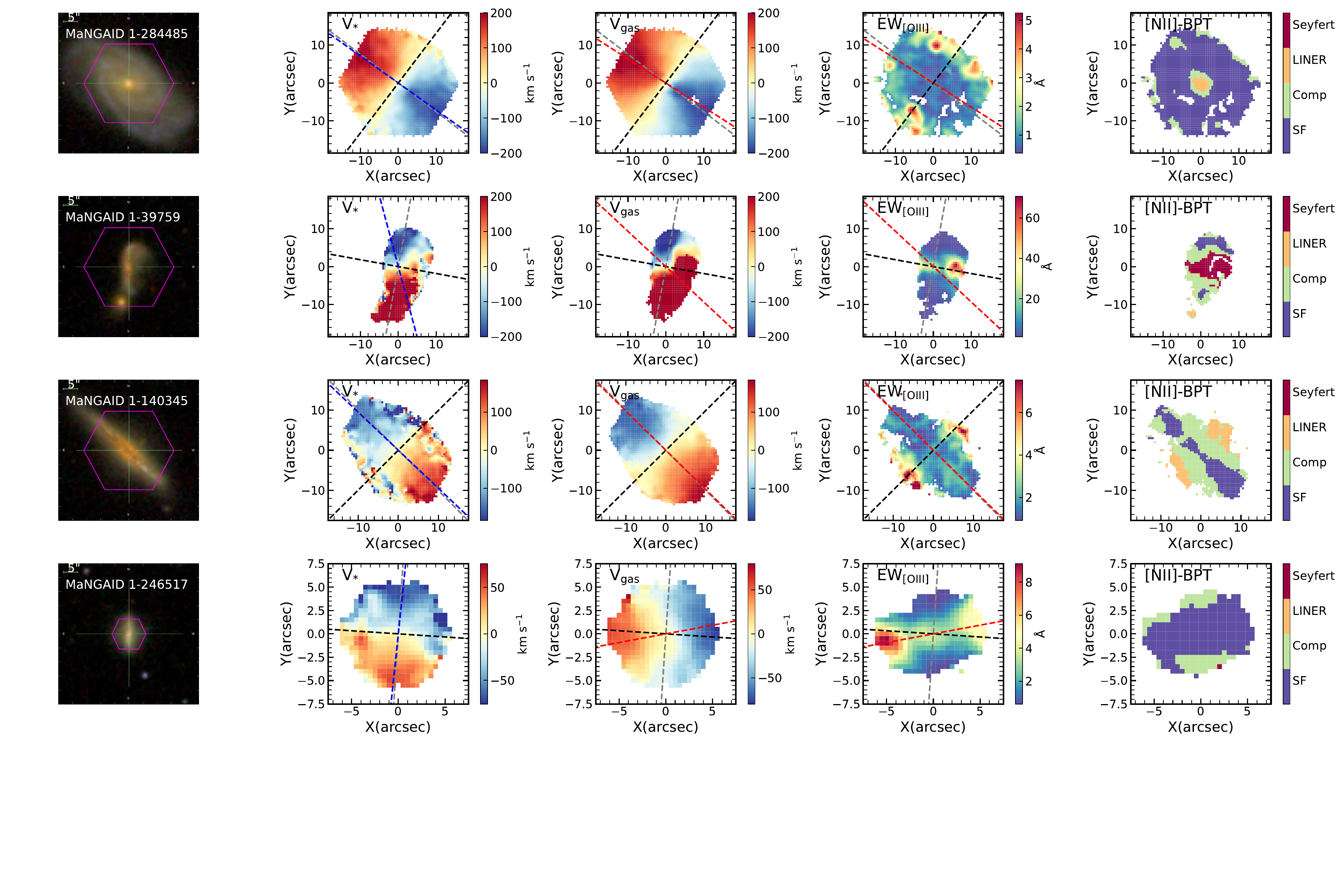}
    \caption{
	Examples of galaxies excluded by eye. 
	The first column shows the SDSS $g, r, i$ color images. The magenta hexagon marks the MaNGA bundle. 
	The second column shows the stellar velocity fields ($\rm V_{*}$). 
	The third column shows the gas velocity fields ($\rm V_{gas}$) traced by {\ha}. 
	In both stellar and gas velocity fields, red side moves away from us, while blue side approaches us. 
	The fourth column shows the {\oiii}$\rm \lambda$5007 equivalent width maps. 
	The fifth column shows the spatially resolved {\nii}-BPT maps. 
	The first row shows a galaxy with $\rm EW_{[OIII]}$ enhanced along minor axis caused by star formation in spiral arms. 
	The second row shows an ongoing merger galaxy. 
	The third row shows a galaxy with $\rm EW_{[OIII]}$ only enhanced at outskirts due to the diffused ionized gas. 
	The last row shows a gas-star misaligned galaxy whose $\rm EW_{[OIII]}$ enhancement aligns with gas kinematic major axis. 
	The black (gray) dashed line in each panel marks the photometric minor (major) axis. 
	The blue and red dashed line in each panel mark the position angle fitted by {\tt PAFIT} package in {\tt PYTHON} \citep{2006MNRAS.366..787K} of stellar and gas components, respectively. 
	}
    \label{fig:del_galaxy}
\end{figure*}

Our selection method is constrained by several criteria, including the galaxy inclination and parameters output by the double Gaussian fitting process. 
We summarize the sample selection criteria in the following: 
\begin{itemize}
    \item [1)] $\rm EW_{[OIII]}$ > 3\AA\ for at least 90\% of spaxels with {\oiii}$\rm \lambda$5007 $\rm S/N>3$. 
	This ensures the selected galaxies have strong enough {\oiii}$\rm \lambda$5007 emission. 
	After taking this criterion, 5870 sources are left. 

    \item [2)] $i \geqslant \ang{60}$ or $i = -\ang{200}$. 
    $i \geqslant \ang{60}$ cut selects edge-on galaxies where biconical cones and galactic disk are spatially separated from each other in the projected sky plane. 
	We keep galaxies with $i = -\ang{200}$ (elliptical galaxies with $\rm fracDeV \geqslant 0.8$) to avoid missing potential biconical galaxies. 
	3698 galaxies are retained in this step. 

    \item [3)] $\lvert \phi_{1} - \ang{90}\rvert \leqslant \phi_{\rm crit1}$, $\lvert \phi_2 - \ang{270}\rvert \leqslant \phi_{\rm crit1}$, where $\phi_{\rm crit1} = \ang{40}$. 
	This criterion ensures the biconical structures along the minor axis of the galaxies. 
	The tolerant range of $\phi_{\rm crit1} = \ang{40}$ comes from empirical estimation. 
	We also visually selected a biconical sample (147 galaxies) from 7577 emission-line galaxies. 
	We apply this cut on $\rm \phi_{1}$ and $\rm \phi_{2}$ to the visually selected sample. 
	Fig.~\ref{fig:criteria}(a) shows how the number of biconical galaxies selected by eye changes with the critical cut $\phi_{\rm crit1}$. 
	It is clear that $\sim 95\%$ of biconical galaxies selected by eyes follow this criterion if we set $\phi_{\rm crit1}>\ang{40}$. 
	For $\phi_{\rm crit1}<\ang{40}$, most biconical galaxies are excluded by this criterion, thus we finally set $\phi_{\rm crit1} = \ang{40}$. 
	After applying the criterion, 1101 galaxies left for further analysis. 

    \item [4)] $\lvert(\phi_{2} - \phi_{1}) - \ang{180} \rvert \leqslant \phi_{\rm crit2}$, where $\phi_{\rm crit2} = \ang{35}$. 
    This guarantees that the two cones of a galaxy are approximately collinear, avoiding galaxies where the two cones are too close to (far-away from) each other. 
	Similar to 3), the $\phi_{\rm crit2} = \ang{35}$ dispersion range comes from empirical estimation. 
	Fig.~\ref{fig:criteria}(b) shows how the number of biconical galaxies selected by eye changes with $\phi_{\rm crit2}$ once we apply this selection criterion. 
	More than 95\% of biconical galaxies selected by eyes follow this cut for $\phi_{\rm crit2}>\ang{35}$. For $\phi_{\rm crit2}<\ang{35}$, the number decreases rapidly. 
	For this reason, we set $\phi_{\rm crit2} = \ang{35}$. 
	Based on this criterion, 897 candidates are left. 

    \item [5)] $ \ang{10} \leqslant \sigma_{1} \leqslant \ang{55}$, $ \ang{10} \leqslant \sigma_{2} \leqslant \ang{55}$. 
	Fig.~\ref{fig:criteria}(c) shows that the distribution of $\sigma_{1}$, $\sigma_{2}$ for biconical galaxies selected by eye. 
	It is clear that the distribution of $\sigma_{1}$, $\sigma_{2}$ ranges from $\ang{10} - \ang{55}$ for galaxies with biconical structures. 
	A high $\sigma$ value indicates less obvious biconical component, while a low $\sigma$ could be due to the outlier data points of a galaxy. 
	We limit $\sigma$ in the range of $\ang{10} - \ang{55}$ to ensure that the biconical profile of $\rm EW_{[OIII]}$ is obvious and real. 
	592 galaxies are left as biconical ionized candidates.
\end{itemize}

We visually inspected these 592 candidates to exclude contamination, 
including 224 galaxies whose $\rm EW_{[OIII]}$ is enhanced by star formation in the spiral arms along minor axis (see the first row in Fig.~\ref{fig:del_galaxy} as an example), 18 ongoing mergers (the second row in Fig.~\ref{fig:del_galaxy}), 
101 galaxies that $\rm EW_{[OIII]}$ only enhanced at outskirts due to the diffused ionized gas (the thrid row in Fig.~\ref{fig:del_galaxy}), 
107 galaxies in which gas and stars are rotating perpendicularly to each other and the gas kinematic major axis aligns with the direction of the $\rm EW_{[OIII]}$ enhanced regions (i.e., the photometric minor axis; the last row in Fig.~\ref{fig:del_galaxy}). 
It is worth noticing that this type of galaxies have similar features as red geysers \citep{2016Natur.533..504C},  
i.e. they are gas-star misaligned galaxies and have bisymmetric enhancement in ionized gas (traced by {\ha} and {\oiii}$\lambda$5007) along the gas kinematic major axis. 
Red geysers are quiescent early-type galaxies which are believed to have galactic-scale outflows driven by AGN activities \citep{2016Natur.533..504C,2018ApJ...869..117R,2021ApJ...913...33R,2021ApJ...922..230R}. 
We exclude these 107 galaxies from our analysis since we can not figure out whether this ionized gas emission enhancement is due to external gas acquisition which increase the gas density along a certain direction (gas kinematic major axis in our cases). 
From the left to right, Fig.~\ref{fig:del_galaxy} shows the SDSS $g, r, i$ color images, the stellar velocity fields, the gas velocity fields, the {\oiii}$\lambda$5007 equivalent width maps, and the spatial resolved {\nii}-BPT maps. 
Finally, 142 biconical ionized galaxies are selected, including 113 edge-on disc galaxies and 29 elliptical galaxies, 
which correspond to $\sim 5\%$ of edge-on emission-line galaxies and $\sim 1\%$ of emission-line elliptical galaxies, respectively.

\subsection{Control samples} \label{control sample}

In order to understand the driving mechanisms of biconical galaxies, we build non-biconical control samples for star-forming and Seyfert biconical galaxies, 
which is selected in the following way: 
\begin{itemize}
    \item [1)] 
	The line ratios measured from the stacked spectra are required to fall within the same ionization region in the {\nii}-BPT diagram \citep{1981PASP...93....5B} as that of biconical galaxies. 
	This criterion ensures that the central regions of both the biconical galaxies and their control samples are ionized by the same mechanism. 
	We stack the spectra of square spaxels (typical spaxel size $\rm \sim 0.3\times0.3kpc^{2}$ for median redshift 0.03 of MaNGA sample) following the method of \citet{2023A&A...674A..85A} within a circular aperture of central 1 kpc radius for all the emission-line galaxies. 
	We apply weighted averages such that if a square spaxel partially contributes to a circular aperture, we weigh the pixel based on the fraction of its enclosed area. 
	\citet{2023A&A...674A..85A} analyze the influence of aperture sizes on AGN selection, 
	suggesting that the central 1 kpc circular region is a good choice for classifying the central ionization state. 

    \item [2)] $ \rvert \Delta i \rvert \leqslant \ang{5}$. 
	To avoid the influence of the inclination, we select control galaxies with similar inclination angles.

    \item [3)] $ \rvert \Delta \rm{log} \textit{M}_{*} \rvert \leqslant 0.1 $. 
    Stellar mass is the most fundamental property of a galaxy and tightly correlates with many other physical parameters. 
		
    \item [4)] $ \rvert \Delta \rm{D}$$_{n}4000 \rvert \leqslant 0.05 $. 
	The global $\rm D$$_{n}4000$ is a reliable indicator of stellar populations. By constraining control galaxies to have similar $\rm D$$_{n}4000$, we ensure they have similar stellar populations. 
			
\end{itemize}

For each biconical galaxy whose central 1 kpc circular regions show Seyfert ionization state, 
we select one control galaxy since the limit number ($\rm \sim 250$ in total) of emission-line Seyfert galaxies in the MaNGA survey. 
Five control galaxies are selected for each star-forming biconical galaxy. 

\section{Results}

\subsection{Morphologies of Biconical Structures}

\begin{figure}
	\centering
	\includegraphics[width=\columnwidth]{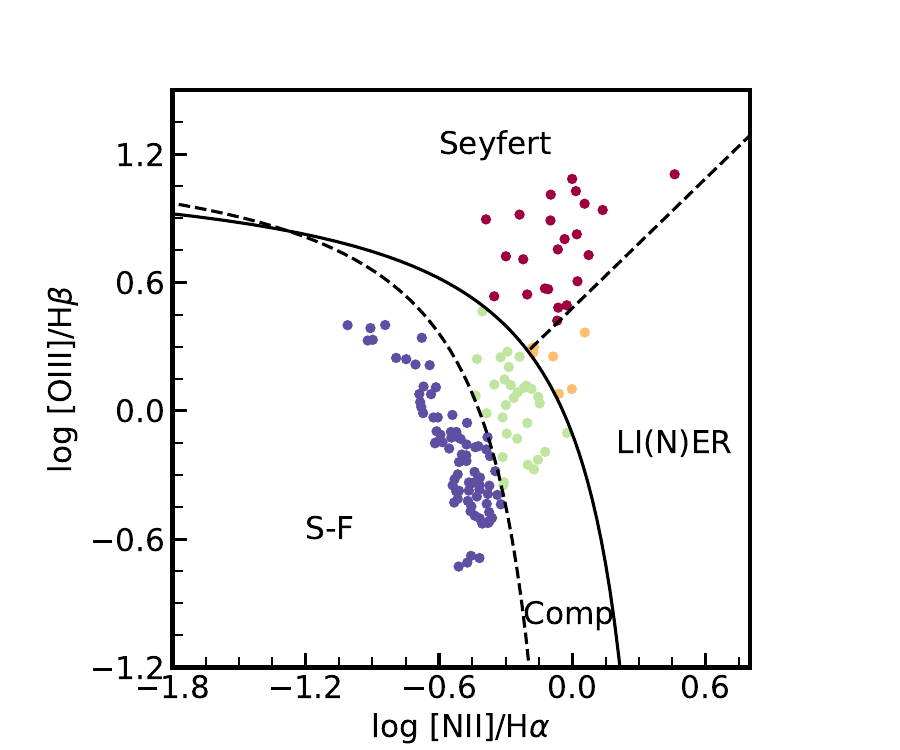}
    \caption{
	{\nii}-BPT diagnosis on stacked spectra measured from a central circular region of 1 kpc radius for galaxies with biconical structures. 
	The solid curve \citep{Kewley_2001} and the dashed curve \citep{2003MNRAS.346.1055K} are the demarcations which separate galaxies into AGNs, composite (green dots), and star-forming (blue dots).
	The Seyferts (red dots) and LI(N)ERs (orange dots) are separated by the dashed straight line \citep{2010MNRAS.403.1036C}.
	}
    \label{fig:sample_center_bpt.pdf}
\end{figure}

\begin{figure*}
	\centering
	\includegraphics[width=2\columnwidth]{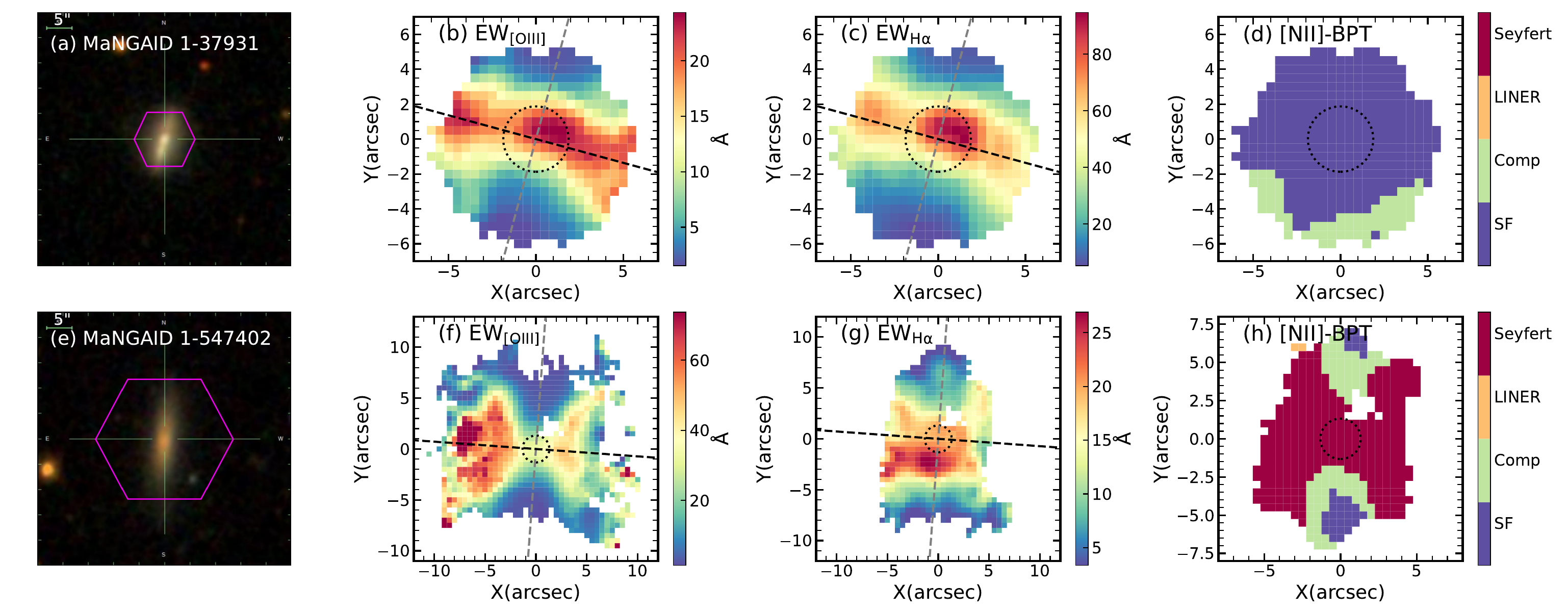}
    \caption{Examples of galaxies with biconical ionized structures. 
	The first row is an example of the star-forming bicone, while the second row is an example of Seyfert bicone. 
	The first column shows the SDSS $g, r, i$ color images of the galaxies. The magenta hexagon marks the MaNGA bundle. 
	The second column shows the {\oiii}$\rm \lambda$5007 equivalent width maps. 
	The third column shows the {\ha} equivalent width maps. 
	The black (gray) dashed line shows the photometric minor (major) axis of the galaxy. 
	The fourth column shows the spatially resolved {\nii}-BPT maps. 
	The different colors represent the different ionization states identified by the {\nii}-BPT diagram. 
	The blue, green, orange, and red spaxels represent the ionization states of star-forming, composite, LI(N)ER, and Seyfert, respectively. 
	Black dashed circle in each map represents aperture of a 1 kpc radius. 
	}
    \label{fig:example_bicone}
\end{figure*}

\begin{figure*}
	\includegraphics[width=2\columnwidth]{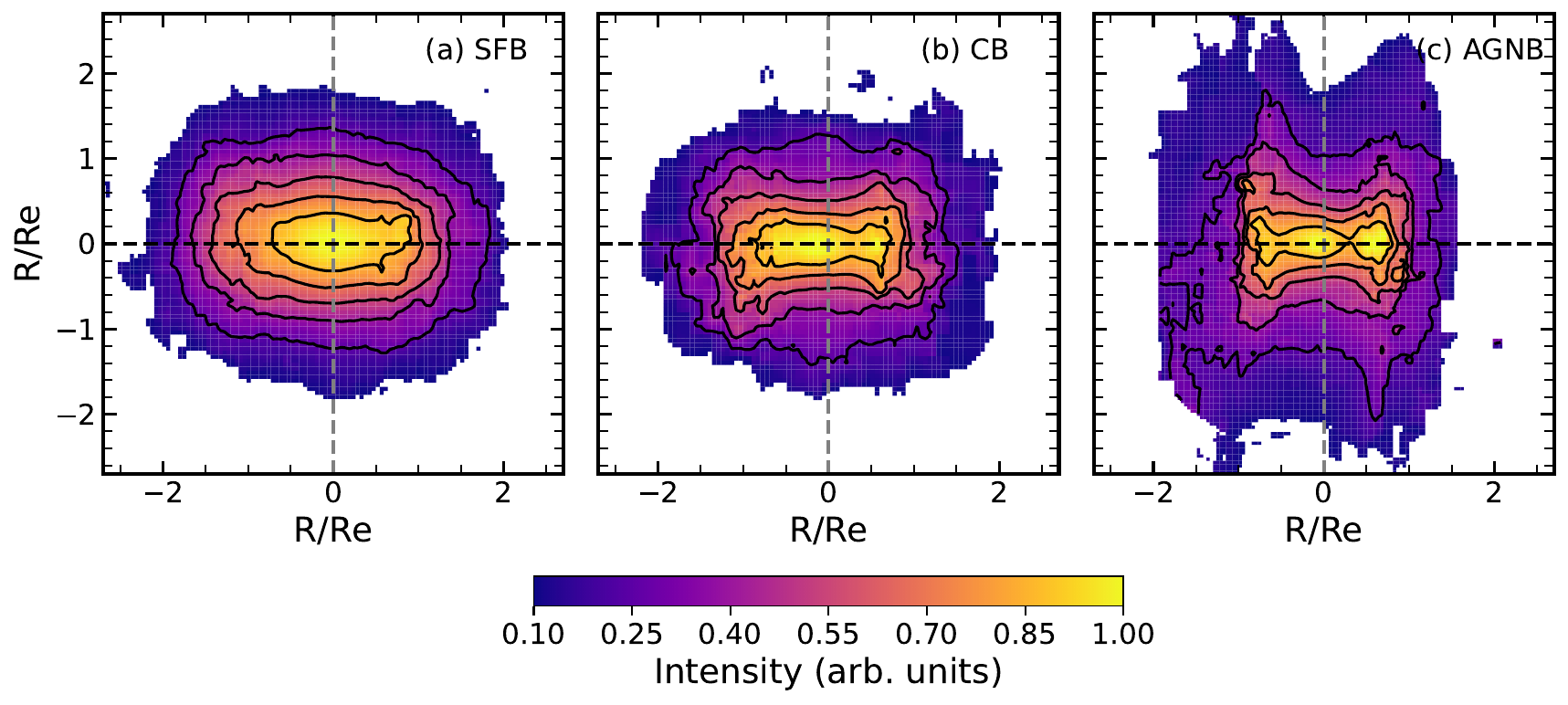}
    \caption{Stacked $\rm EW_{[OIII]}$ maps. 
	Left column for star-forming bicones, middle column for composite bicones, and right column for AGN bicones. 
	The color bar shows the normalized intensity of equivalent width. 
	The black (gray) dashed line marks the photometric minor (major) axis. 
	}
    \label{fig:bicone_morphology}
\end{figure*}

Biconical structures can be triggered by both stellar wind from massive stars and SNe \citep{1985Natur.317...44C, 1990ApJS...74..833H}, as well as central SMBH activities \citep{doi:10.1146/annurev-astro-082214-122316,2019MNRAS.490.3025R}. 
In order to distinguish the main driving mechanism of biconical structures, 
we divide biconical galaxies into several subgroups according to their ionization states of the central regions. 

The widely used system for spectral classification of emission-line galaxies is the diagnostic diagrams suggested by \citet*[][generally referred as BPT diagrams]{1981PASP...93....5B}. 
The standard BPT diagram rely on emission line ratios $\rm {\oiii}\lambda5007/{\hb}$ vs. $\rm {\nii}\lambda6583/{\ha}$, $\rm {\sii}\lambda \lambda6717,31/{\ha}$ or $\rm {\oi}\lambda6583/{\ha}$. 
In this work, we apply the {\nii}-BPT diagram to the stacked spectra for a central circular region of 1 kpc radius. 
Fig.~\ref{fig:sample_center_bpt.pdf} show the classification of biconical galaxies based on ionization states of the central regions. 
The solid curve \citep{Kewley_2001} and the dashed curve \citep{2003MNRAS.346.1055K} represent the demarcation lines that separate the AGN, composite, and star-forming galaxies regions. 
The dashed straight line \citep{2010MNRAS.403.1036C} distinguishes Seyferts from Low-Ionization (Nuclear) Emission-line Regions (LI(N)ERs). 
According to the {\nii}-BPT diagram measured from stacked spectra of a central circular region of 1 kpc radius, 
we divided the biconical galaxies into three subgroups: 81 star-forming bicones (SFBs), 30 AGN bicones (AGNBs, composed of 23 Seyferts and 7 LI(N)ERs), and 31 composite bicones (CBs).

Fig.~\ref{fig:example_bicone} shows examples of a SFB (top panel) and a Seyfert bicone (bottom panel). 
The first column shows the SDSS $g,r,i$-band images. 
The second column shows {\oiii}$\lambda5007$ equivalent width maps. It is obvious that there is a significant $\rm EW_{[OIII]}$ enhancement along the photometric minor axis (black dashed line). 
The third column shows $\rm EW_{H\alpha}$ maps. 
Although $\rm EW_{H\alpha}$ is also enhanced along minor aixs, it is not as obvious as $\rm EW_{[OIII]}$, which is not only for Seyfert bicones, but also for star-forming bicones. 
The fourth column shows the spatially resolved {\nii}-BPT diagrams, and the blue, green, orange, and red spaxels represent the ionization states of star-forming, composite, LI(N)ER, and Seyfert, respectively. 
We go through the $\rm EW_{[OIII]}$ map of all the biconical galaxies, finding that galaxies with different central ionization states show different biconical morphologies. 
$65\%$ Seyfert galaxies have obvious hourglass biconical shapes (see Fig.~\ref{fig:example_bicone}(f)), very thin at the center and becomes wider to the outsides along minor axes. 
The star-forming galaxies do not have such obvious difference between the central and outside regions (see Fig.~\ref{fig:example_bicone}(b)). 

In order to have an idea about the difference between biconical shapes in different kinds of galaxies statistically, we stack the $\rm EW_{[OIII]}$ maps for SFBs, CBs and AGNBs, respectively. 
First, we normalize the $\rm EW_{[OIII]}$ maps by central intensity to set the peak value to 1, ensuring consistency in the intensity scaling. 
Second, we rotate the $\rm EW_{[OIII]}$ maps, putting the photometric major axis in the vertical direction. 
Finally, we stack the normalized and rotated $\rm EW_{[OIII]}$ maps of galaxies with the same central ionization state. 
Fig.~\ref{fig:bicone_morphology} shows stacked $\rm EW_{[OIII]}$ maps of SFBs, CBs, and AGNBs, respectively. 
As expected, the morphologies of biconical structures are different between SFBs, CBs, and AGNBs, with a hourglass shape for AGNBs and a bar shape for SFBs. 
The morphology of CBs falls in-between SFBs and AGNBs. 

In star-forming galaxies, galactic winds are triggered by multiple superbubbles, which are generated by the explosion of SNe in various regions within the galactic disk \citep{1980A&A....87..152H,2003RMxAC..17...47H,2022A&A...659A.153R,2023A&A...678A..84D}. 
Therefore, it is expected the base of the biconical structure in SFBs is extended along the galactic disk. 
However, AGN-driven galactic winds are typically powered by the central SMBHs, thus the base of the biconical structure is point-like, and the ionized gas becomes more dispersed at the outskirts \citep{2005ARA&A..43..769V,2014ApJ...797...90S,Tanner2022AGNoutflow}. 
As a result, the biconical ionized structure generated by AGN is expected to resemble an hourglass shape. 
Recent observations of local dwarf galaxies \citep{2023AAS...24136109A} also reveal that outflows triggered by AGN and stellar activities exhibit different morphologies. 
In the case of composite biconical ionized galaxies, both black hole activity and star formation contribute to the formation of biconical ionized structures, 
thus the morphologies of ionization cones are the transitional states between SFBs and AGNBs.

\subsection{Primary Driver of Biconical Structures}
\subsubsection{Biconical Ionized Structures in Star-Forming Galaxies}

\begin{figure*}
	\centering
	\includegraphics[width=2\columnwidth]{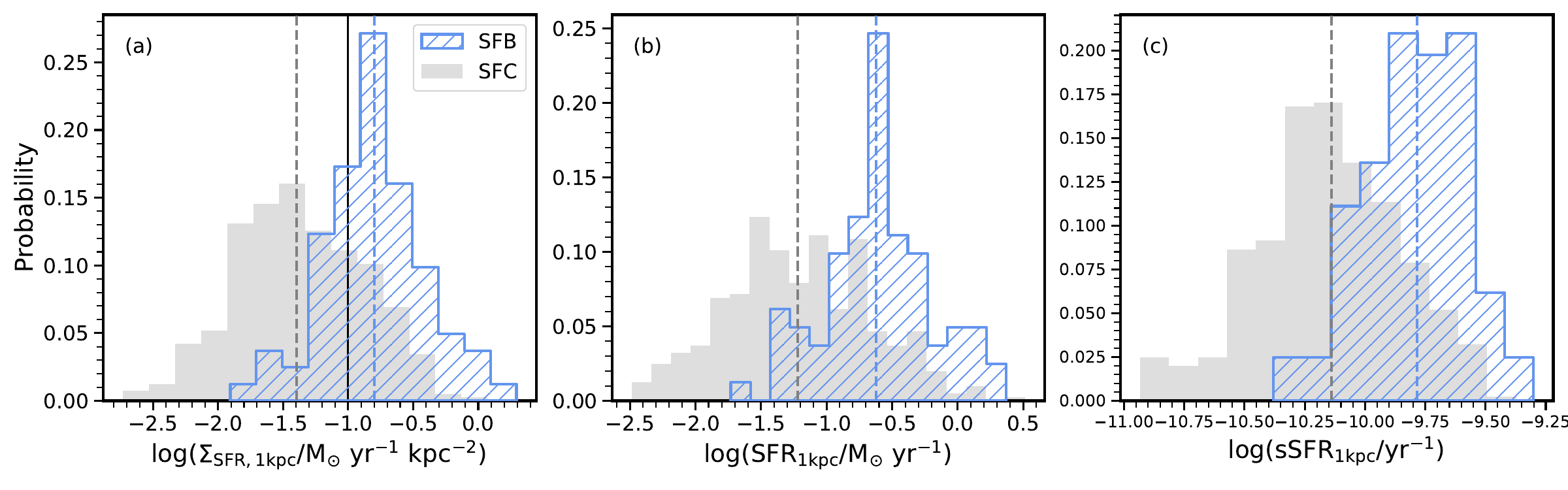}
    \caption{
	The statistical distributions of physical parameters of SFBs and their control samples. 
	(a) SFR surface density within central 1 kpc. Black solid line marks the threshold of {\sigmasfr} $= \rm 0.1\ M_{\odot}\ yr ^{-1}\ kpc^{-2}$. 
	(b) SFRs within central 1 kpc. 
	(c) Specific SFRs within central 1 kpc. 
	The blue histograms represent star-forming bicones while the gray histograms represent their control samples.
	The blue and gray dashed vertical lines in each panel indicate the median value of star-forming bicones and their control samples, respectively. 
	Note when estimating these three physical parameters, we only use star-forming spaxels diagnosed by {\nii}-BPT diagram within the central circular aperture of a 1 kpc.  
}
    \label{fig:sf_bicone_hist}
\end{figure*}

In star-forming galaxies, the biconical ionized structures are believed to be triggered by star-formation activities. 
The parameters that used to describes the strength of star-formation activities include star formation rate (SFR), the surface density of SFR ({\sigmasfr}), as well as specific SFR (sSFR $\equiv {\rm SFR}/M_{*}$). 
\citet{2002ASPC..254..292H} pointed out the importance of the central {\sigmasfr} in governing biconical outflows in star-forming galaxies. 
Based on a single fiber spectral sample of star-forming galaxies in SDSS DR7, \citet{Chen_2010} study outflow using NaD absorption lines, finding that for the outflow component, its $\rm EW_{NaD}$ primarily depends on {\sigmasfr} 
and the outflow velocities ($V_{\rm eff}$) dose not depends strongly on any physical parameters except a shallow trend with {\sigmasfr} ($V_{\rm eff} \propto$ {\sigmasfr}$^{0.1}$). 
\citet{2022MNRAS.516.3092B} find a batch of face-on biconical outflow galaxies driven by star-formation in MaNGA, which also show a similar dependence ($V_{\rm eff} \propto$ {\sigmasfr}$^{0.2\sim0.3}$). 
In this section, we compare the distribution of {\sigmasfr}, SFR and sSFR integrated over star-forming spaxels within the circular aperture of 1 kpc radius between galaxies with ionized bicones and their control samples without bicones. 

We estimate the dust attenuation corrected flux as 
\begin{equation*}
	F_{\rm{int}}(\lambda) = F_{\rm{obs}}(\lambda)\times 10^{-0.4 k(\lambda) E(B-V)}, 
\end{equation*}
where $F_{\rm{int}}(\lambda)$ is the intrinsic flux, and $F_{\rm{obs}}(\lambda)$ represents observed flux. 
The color excess $E(B-V) = 0.934 \times {\rm ln}[(F_{\rm H\alpha}/F_{\rm H\beta})/2.86]$, and $k(\rm \lambda)$ is dust attenuation curve \citep{Calzetti_2000}. 
SFR is derived from dust attenuation corrected $\rm {\ha}$ luminosity \citep{doi:10.1146/annurev.astro.36.1.189} of star-forming spaxels as 
\begin{equation*}
	\rm{SFR} (\textit{M}_{\rm\odot}\ yr ^{-1}) = 7.9 \times 10^{-42}\textit{L}_{\rm H\alpha}\rm{(erg\ s^{-1})}. 
\end{equation*}
The sSFR is estimated as ${\rm SFR}/M_{*}$, where $M_{*}$ is spatial resolved stellar mass taken from {\tt PIPE3D} value-added catalog \citep{2016RMxAA..52..171S}. 
{\sigmasfr} is estimated as ${\rm SFR}/A$, where $A$ is the projected area of the star-forming regions in unit of $\rm kpc^{2}$. 

Fig.~\ref{fig:sf_bicone_hist} shows distributions of {\sigmasfrkpc}, {\sfrkpc} and {\ssfrkpc}, respectively. 
Note when estimating these three physical parameters, we only use star-forming spaxels diagnosed by {\nii}-BPT diagram within the central circular aperture of a 1 kpc. 
The blue histograms represent SFBs, while the gray shadow histograms represent the star-forming control samples (SFCs). 
The blue (gray) vertical lines represent the median value of parameters for SFBs (SFCs), and black vertical line marks the threshold of $\rm \Sigma_{SFR} = 0.1\ M_{\odot}\ yr ^{-1}\ kpc^{-2}$ \citep{1980A&A....87..152H}. 
The median values of {\sigmasfrkpc}, {\sfrkpc} and {\ssfrkpc} in SFBs are 0.66 dex, 0.58 dex and 0.38 dex higher than that of SFCs, respectively. 
The distribution of {\sigmasfrkpc} shows the largest difference between SFBs and SFCs. 
We also compare the global SFR, {\sigmasfr} and sSFR histograms between biconical samples and their controls, finding that the difference is much smaller than that of these parameters within 1 kpc. 
These results imply that the level of star-forming activities in the central regions of SFBs is more intense than that observed in SFCs, 
suggesting a strong connection between central star-forming activities and the formation of biconical structures. 

In order to quantify the difference between SFBs and SFCs, 
we perform the two-sample Kolmogorov-Smirnov test \citep[K-S Test;][]{1572824501049496320,1363670321263386240}, 
which is a non-parametric statistical method used to determine whether two independent samples exhibit significant differences in their distributions. 
The null hypothesis is that SFBs and SFCs are drawn randomly from parent samples, and they show no obvious differences in the levels of star-formation activities. 
We adopt 0.01 as a critical $p$-value to reject the null hypothesis. 
In the case of {\sfrkpc} and {\ssfrkpc}, the K-S test results in a low $p$-value of $\sim 10^{-14}$, 
indicating obvious differences in these two parameters between SFBs and SFCs. 
With a remarkably low $p$-value of $\sim 10^{-16}$ for {\sigmasfrkpc}, it strongly suggests that the central {\sigmasfrkpc} is the primary driver of biconical structures in SFBs. 

\subsubsection{Biconical Ionized Structures in Seyfert Galaxies}

\begin{figure*}
	\includegraphics[width=1.0\linewidth]{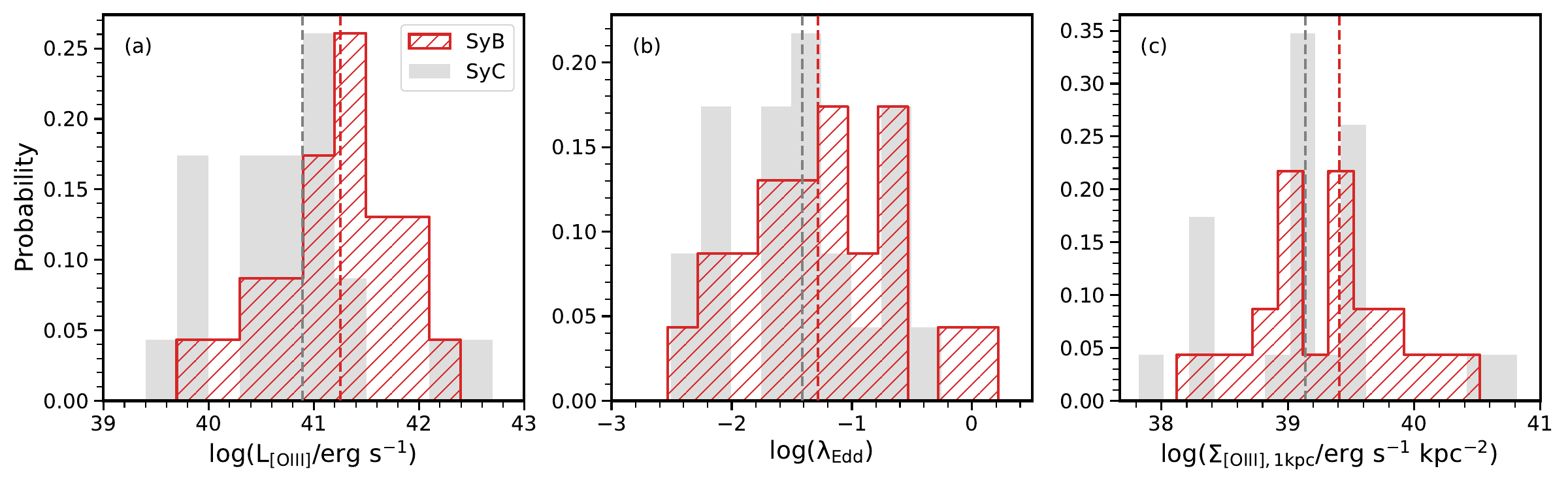}
    \caption{
	The statistical distributions of physical parameters of SyBs and their control samples. 
	(a) {\oiii}$\rm \lambda$5007 luminosities.
	(b) Eddington ratios.
	(c) {\oiii}$\rm \lambda$5007 surface brightness for Seyfert spaxels diagnosed by {\nii}-BPT diagram within the central circular aperture of a 1 kpc radius. 
	The red histograms represent Seyfert biconical galaxies while the gray shadow histograms show their control samples. 
	The red and gray dashed vertical lines in each panel imply the median value of Seyfert biconical galaxies and their control samples, respectively.
	}
    \label{fig:Seyfert_bicone_hist}
\end{figure*}

In Seyfert bicones, the primary ionization source is believed to be the central SMBH. 
The {\oiii}$\rm \lambda$5007 luminosity ({\loiii}), Eddington ratio ({\eddratio}) and surface brightness of {\oiii}$\lambda$5007 emission for Seyfert spaxels within the central circular aperture of 1 kpc radius ({\sigmaoiiikpc}) can serve as indicators of SMBH activity strength. 
In this section, in order to understand the origin and primary driving mechanisms of biconical structures in Seyfert galaxies, 
we compare Seyfert galaxies exhibiting biconical structures (SyBs) with non-biconical Seyfert control galaxies (SyCs).

We estimate the bolometric luminosity as $L_{\rm{bol}} = 700L_{\rm [OIII]}$, 
where $L_{\rm [OIII]}$ is the dust attenuation corrected {\oiii}$\lambda$5007 luminosity \citep{2009ApJ...705..568L}. 
The {\oiii}$\rm \lambda$5007 luminosity is estimated from spaxels falling within the Seyfert regions on the {\nii}-BPT diagram. 
The black hole mass ($M_{\rm BH}$) is estimated from the well-known $M_{\rm BH}-\sigma_{*}$ relation \citep{2000ApJ...539L..13G}, 
where $\sigma_{*}$ is the flux-weighted mean stellar velocity dispersion of all spaxels within 1$R_{\rm e}$ of a galaxy. 
The Eddington ratio is then calculated as $\lambda_{\rm{Edd}} = L_{\rm{bol}}/L_{\rm{Edd}} $, where $L_{\rm{Edd}}$ is the Eddington luminosity.

From left to right, Fig.~\ref{fig:Seyfert_bicone_hist} shows the distributions of {\loiii}, {\eddratio}, and {\sigmaoiiikpc}, respectively. 
The red histograms represent SyBs, while gray shadow histograms represent SyCs. 
The median values of {\loiii} and {\sigmaoiiikpc} of SyBs are about 0.36 dex and 0.27 dex higher than those of SyCs, respectively. 
However, the difference in {\eddratio} between the SyBs and SyCs is not obvious. 
We also apply the K-S test for SyBs and SyCs, resulting in a $p$-value of 0.07, 0.22 and 0.21 for {\loiii}, {\eddratio}, and {\sigmaoiiikpc}, respectively. 
The $p$-values are greater than 0.01 for all statistical parameters, suggesting no significant differences in the distributions of these three parameters between SyBs and SyCs. 

One possible explanation for the lack of differences in the strength of BH activities between SyBs and SyCs is that 
the accretion disk and the galactic disk are not necessarily coplanar \citep{1999ApJ...516...97N,SCHMITT2002231,2012MNRAS.425.1121H}. 
Typically, AGN-driven outflows are generally along the direction perpendicular to the accretion disk \citep{2013LRR....16....1A}. 
Thus, the outflows in Seyferts are not necessarily perpendicular to the galactic disk. 
In the case that the accretion disk and the galactic disk are misaligned, i.e., perpendicular to each other, 
we can not find the biconical ionized structures using the current sample selection method, 
even the AGN activities are comparable to or larger than those of the control sample.

\section{The Formation of Biconical Structure in LI(N)ERs}

\begin{table*}
	\caption{The MaNGA ID, identifications of previous studies, dust attenuation corrected {\oiii}$\lambda$5007 luminosity extracted from LI(N)ER spaxels within the central circular aperture of a 1 kpc radius, global $\rm D$$_{n}4000$, Sérsic index and inclinations of LI(N)ER bicones. }
	\label{tab:LINER_bicone}
	\resizebox{2\columnwidth}{!}{
	\begin{tabular}{cccccc} 
		\hline
		MaNGA ID & Previous Identification  & log$L_{\rm [OIII]}(\rm erg\ s^{-1})$ & Global $\rm D$$_{n}4000$ & Sérsic index & inclination($^{\circ}$) \\ 
		\hline
		1-275456 & Radio-AGN (BH12; C20)  & 40.23 & 1.556 & 1.283 & 69.8   \\
		1-261224 & AGN (B17)              & 39.88 & 1.672 & 3.494 & Elliptical \\
		1-151307 & AGN (B17)              & 39.36 & 1.647 & 1.283 & 76.3   \\
		1-135668 & -                      & 39.40 & 1.713 & 1.360 & 87.0   \\
		1-197780 & -                      & 39.67 & 1.544 & 1.253 & 87.0   \\
		1-153613 & -                      & 39.80 & 1.470 & 1.196 & 81.7   \\
		1-233563 & -                      & 39.86 & 1.539 & 1.668 & 64.2   \\
		\hline
	\end{tabular}
	}
\end{table*}
\begin{figure*} 
	\includegraphics[width=2\columnwidth]{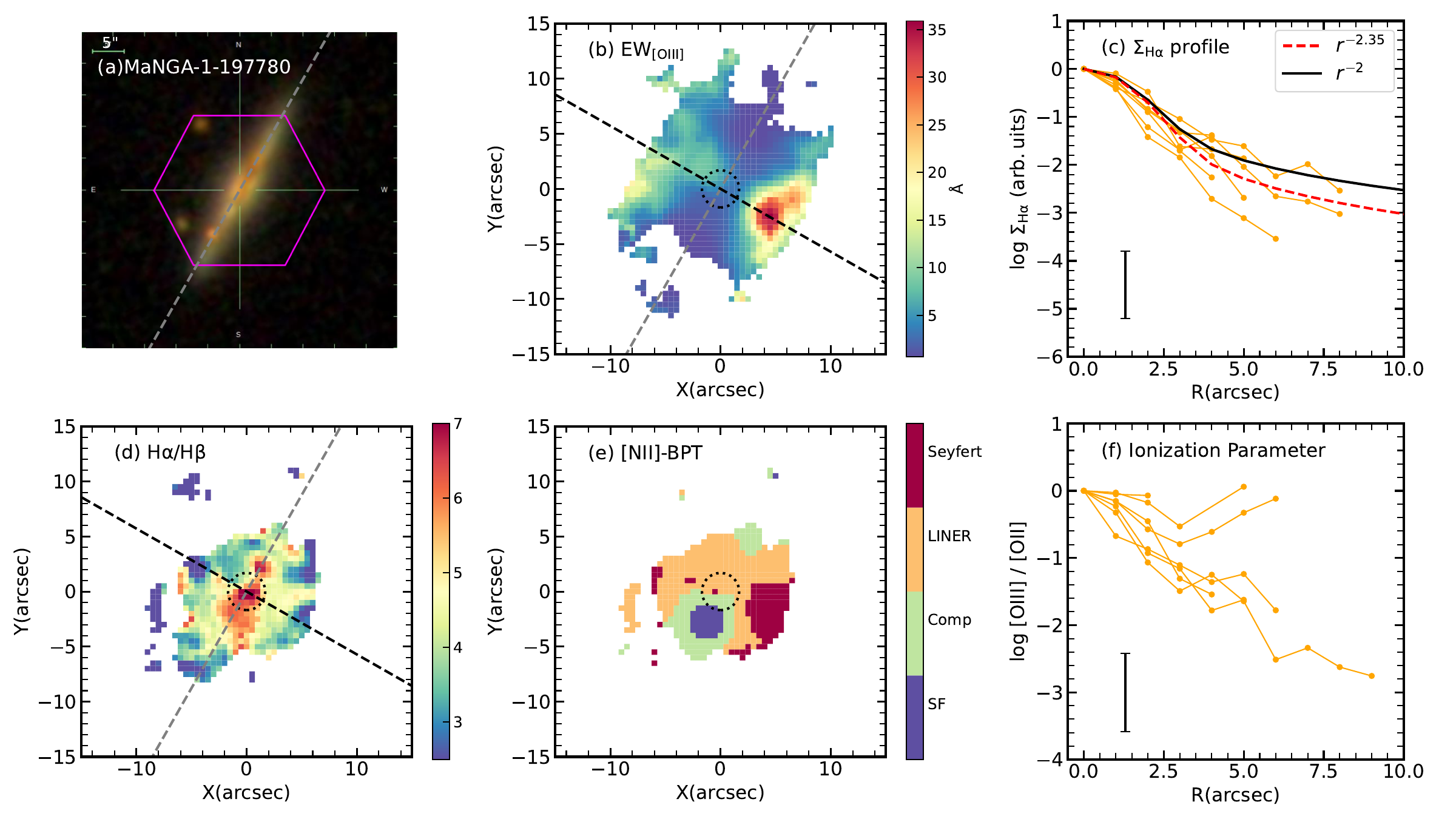} 
    \caption{
		An example of LI(N)ER biconical galaxy. 
	(a) SDSS $g, r, i$ color image of the galaxy. The magenta hexagon marks the MaNGA bundle. 
	(b) {\oiii}$\rm \lambda$5007 equivalent width map. 
		The black (gray) dashed line shows the photometric minor (major) axis. 
	(c) {\sigmaha} radial gradients of all LI(N)ER bicones. 
		The black solid curve represents a $r^{-2}$ model convolved with the median $r$-band PSF ($\rm 2.47\arcsec$) of these 7 LI(N)ER biconical galaxies.  
		The red dashed curve shows a global {\sigmaha} radial gradient of $r^{-2.35}$ for LI(N)ER bicones, 
		which is the median of the $r^{-\alpha}$ models of each LI(N)ER biconical galaxy. 
		The error bar in bottom left shows typical $\rm \pm 1\sigma$ range. 
	(d) $\rm H\alpha/H\beta$ map. 
	(e) Spatially resolved {\nii}-BPT map. 
		The different colors of spaxels represent the different ionization states identified by the {\nii}-BPT diagram. 
	(f) Gradients for gas ionization parameter traced by {\oiii}$\lambda$5007/{\oii}$\lambda\lambda$3727,29 of all LI(N)ER bicones. 
		The error bar in bottom left shows typical $\rm \pm 1\sigma$ range. 
		Black dashed circle in each map represents aperture of a 1 kpc radius. 
	}
	\label{fig:LINER_bicone}
\end{figure*}

The LI(N)ER population was first proposed by \citet{1980A&A....87..152H}. 
Originally thought to be low-luminosity AGNs, the ionization mechanisms of LI(N)ER has been debated. 
Since the development of spatially resolved IFU observations, it is becoming increasingly clear that most galaxies with LI(N)ER regions are unlikely to be low luminosity-AGNs \citep{2010MNRAS.402.2187S,2016MNRAS.461.3111B}. 
The physical size of the LI(N)ER region can extend to the kpc scale \citep[e.g.][]{1983ApJS...52..229K,2006MNRAS.366.1151S}, which is unlikely to be photo-ionized by a nuclear source. 

Through detailed analysis of galaxies with LI(N)ER regions in MaNGA survey, \citet{2016MNRAS.461.3111B} find a flat radial distribution in the ionization parameters (traced by {\oiii}$\lambda$5007/{\oii}$\lambda\lambda$3727,29) of gas 
as well as a shallower than $r^{-2}$ radial gradient in {\ha} surface brightness statistically. 
These two observational results strongly suggest that most LI(N)ER emission is due to diffuse stellar sources (i.e., hot evolved stars) rather than a central point source. 
Shocks also play a role in generating LI(N)ER-like line ratio in the merger systems. 
Considering the energy budget in LI(N)ERs is not AGN or star formation, one natural question is: what is the primary driver of biconical structures in these galaxies? 

Among the biconical galaxies, 7 out of 30 AGNBs are classified as LI(N)ER based on {\nii}-BPT diagnosis diagram. 
According to the classification of \citet{2016MNRAS.461.3111B}, 7 LI(N)ER biconical galaxies in this work are all central LIER (c-LIER). 
For these biconical galaxies with LI(N)ER-like line ratio in the central regions, one is identified as radio-AGN \citep[][hereafter BH12 and C20]{2012MNRAS.421.1569B,2020ApJ...901..159C}. 
Two of them are classified as low-luminosity AGN using {\heii} $\lambda$4685 emission line diagnosis \citep[][B17]{2017MNRAS.466.2879B}. 
Table.~\ref{tab:LINER_bicone} summarizes the properties of the host galaxies with biconical structures for LI(N)ERs, including {\oiii}$\lambda$5007 luminosity extracted from LI(N)ER spaxels within the central circular aperture of a 1 kpc radius, $\rm D$$_{n}4000$, Sérsic index, and inclination. 
Based on these results, we conclude that the hosts are edge-on disk galaxies with {\loiii} $\rm \sim10^{40}erg\ s^{-1}$ extracted from only LI(N)ER spaxels within the central circular aperture of a 1 kpc radius, suggesting a relatively strong activity of the central black hole. 
The global $\rm D$$_{n}4000$ of these galaxies ranges from $1.5-1.7$, 
implying that they hold the transition stellar populations between star-forming galaxies ($\rm D$$_{n}4000 < 1.5$) and quiescent galaxies ($\rm D$$_{n}4000 > 1.7$).

Fig.~\ref{fig:LINER_bicone} illustrates an example of LI(N)ER biconical galaxies. 
Fig.~\ref{fig:LINER_bicone}(a) shows the SDSS $g, r, i$ color image of the galaxy. 
Fig.~\ref{fig:LINER_bicone}(b) shows the $\rm EW_{[OIII]}$ map, where the black (grey) dashed line marks the minor (major) axis of galaxy and black dashed circle represents aperture of a 1 kpc radius. 
We can clearly see the biconical enhanced structure along the minor axis in $\rm EW_{[OIII]}$ map. 
Fig.~\ref{fig:LINER_bicone}(c) shows dust attenuation corrected {\sigmaha} radial gradients for all LI(N)ER biconical galaxies, where each orange line represents a galaxy, and error bar in bottom left shows the typical $\rm \pm 1\sigma$ range. 
We fit the radial gradient of dust attenuation corrected {\sigmaha} using $r^{-\alpha}$ model convolved with the $r$-band PSF for each galaxy. 
The red dashed curve with a radial gradient of $r^{-2.35}$ in Fig.\ref{fig:LINER_bicone}(c) shows the median of the $r^{-\alpha}$ models of each LI(N)ER biconical galaxy. 
For companion, we also show $r^{-2}$ model convolved with the median $r$-band PSF ($\rm 2.47\arcsec$) of these 7 LI(N)ER biconical galaxies in black solid curve. 
The $r^{-2.35}$ radial gradient is roughly consistent with $r^{-2}$ model which is predicted by the photoionization of a central point-like source. 
Fig.~\ref{fig:LINER_bicone}(d) shows the $\rm H\alpha/H\beta$ map of one LI(N)ER biconical galaxies, which is a good tracer for dust attenuation.  
We can clearly see there is a strong dust attenuation traced by high values of $\rm H\alpha/H\beta$ along the major around central regions. 
Fig.~\ref{fig:LINER_bicone}(e) shows the spatially resolved {\nii}-BPT map, where the blue, green, orange and red spaxels represent the SF, composite, LI(N)ER and Seyfert regions identified by the {\nii}-BPT diagram, respectively. 
Fig.~\ref{fig:LINER_bicone}(f) shows the ionization parameter gradients of gas measured by the flux ratio of dust attenuation corrected {\oiii}$\lambda$5007 and {\oii}$\lambda\lambda$3727,29 for these 7 LI(N)ER biconical galaxies, where each orange line represents a galaxy, and error bar in bottom left shows typical $\rm \pm 1\sigma$ range. 
The ionization parameter gradients reveal an overall negative gradient, at least for regions within $2.5\arcsec$ whose size is consistent with the median $r$-band PSF, indicating emission from a central point source rather than extended source. 

We suggest the LI(N)ER bicones are driven by obscured AGNs or AGN echoes for the following reasons:
(a) The median value of dust attenuation corrected {\oiii}$\lambda$5007 luminosity extracted from LI(N)ER spaxels within the central 1 kpc circular radius of LI(N)ER biconical galaxies is $\rm \sim10^{40}erg\ s^{-1}$, indicating a relatively active black hole; 
(b) There is strong dust attenuation traced by $\rm H\alpha/H\beta$ along the major around central regions of LI(N)ER biconical galaxies; 
(c) The {\ha} surface brightness have a radial gradients of $\sim r^{-2.35}$, indicating emission from a central point-like source; 
(d) The gas ionization parameters decreased from the center to at least $2.5\arcsec$, also indicating emission from a central point source rather than extended source.

\section{Conclusions}

In this work, based on the MaNGA IFU observation, we developed a new method to select galaxies with ionized bicones, finding 142 bicones from 7577 emission-line galaxies in MaNGA survey.
We separate these 142 galaxies into 81 star-forming galaxies, 31 composite galaxies, and 30 AGNs based on {\nii}-BPT diagnosis diagram. 
We study the morphologies and primary driving mechanisms of these bicones. 
The main conclusions include: 

\begin{itemize}

	\item [1)]
		The biconical morphologies vary with the central ionization states. 
		AGN bicones show hourglass structures while the SF bicones have bar-like structures. 
		Due to the combined effects of central black hole and star-formation activities, the biconical structures in the composite galaxies have a transitional morphology between SFBs and AGNBs. 

    \item [2)] 
		SFBs have more intense star-formation activities in their central regions: The {\sigmasfrkpc}, {\sfrkpc}, and {\ssfrkpc} of SFBs are higher than their non-biconical control samples. 
		The most significant difference is in {\sigmasfrkpc}, which strongly suggests that the {\sigmasfr} of the central region primarily drives the biconical structures in SFBs. 
		
    \item [3)] 
		The dust attenuation corrected {\oiii}$\lambda$5007 luminosity, Eddington ratio and central 1 kpc {\oiii}$\lambda$5007 surface brightness extracted from Seyfert spaxels show no significant difference between Seyfert biconical galaxies and their non-biconical control galaxies. 
		The lack of difference in the strength of black hole activities between SyBs and SyCs can be explained as the accretion disk and the galactic disk are not necessarily coplanar. 

	\item [4)] 
		The host of LI(N)ER bicones are dominated by edge-on disk galaxies, and they have intermediate stellar populations with $\rm 1.5< D$$_{n}4000 < 1.7$.  
		The median value of dust attenuation corrected {\oiii}$\lambda$5007 luminosity of LI(N)ER biconical galaxies extracted from LI(N)ER spaxels within central 1 kpc radius is $\rm \sim10^{40}erg\ s^{-1}$. 
		The high {\ha}/{\hb} values at the center indicate strong dust attenuation in the host galaxies of LI(N)ER bicones. 
		Both the {\ha} surface brightness radial gradients and the ionization parameter decreases with increasing radius, both indicating a central point-like ionization source, 
		These observational results suggest that the LI(N)ER biconical galaxies are triggered by obscured AGNs or AGN echoes. 

		The biconical galaxy sample in this work are outflowing candidates. 
		Detailed kinematic analysis will help us to further confirm whether there are outflows in these galaxies, which will be carried out in the future works. 

\end{itemize}

\section*{Acknowledgements}

We thank the anonymous referee for thoughtful and constructive comments that improved this paper.
Y.M.C. acknowledges support from the National Natural Science Foundation of China (NSFC grants 12333002, 11573013, 11733002, 11922302), 
the China Manned Space Project with NO. CMS-CSST-2021-A05. 
This work was supported by the research grants from the China Manned Space Project, the second-stage CSST science project ‘Investigation of small-scale structures in galaxies and forecasting of observations’.
DB is partly supported by RSCF grant 22-12-00080. 

Funding for the Sloan Digital Sky Survey IV has been provided by the Alfred P.
Sloan Foundation, the U.S. Department of Energy Office of Science, and the Participating Institutions.
SDSS-IV acknowledges support and resources from the Center for High-Performance Computing at
the University of Utah. The SDSS website is www.sdss.org.

SDSS-IV is managed by the Astrophysical Research Consortium for the Participating Institutions of the SDSS Collaboration 
including the Brazilian Participation Group, the Carnegie Institution for Science, Carnegie Mellon University, the Chilean Participation 
Group, the French Participation Group, Harvard-Smithsonian Center for Astrophysics, Instituto de Astrof\'isica de Canarias, The Johns 
Hopkins University, Kavli Institute for the Physics and Mathematics of the Universe (IPMU) / University of Tokyo, Lawrence Berkeley 
National Laboratory, Leibniz Institut f\"ur Astrophysik Potsdam (AIP), Max-Planck-Institut f\"ur Astronomie (MPIA Heidelberg), 
Max-Planck-Institut f\"ur Astrophysik (MPA Garching), Max-Planck-Institut f\"ur Extraterrestrische Physik (MPE), National 
Astronomical Observatories of China, New Mexico State University, New York University, University of Notre Dame, 
Observat\'ario Nacional / MCTI, The Ohio State University, Pennsylvania State University, Shanghai Astronomical Observatory, 
United Kingdom Participation Group, Universidad Nacional Aut\'onoma de M\'exico, University of Arizona, University of Colorado Boulder, 
University of Oxford, University of Portsmouth, University of Utah, University of Virginia, University of Washington, University of Wisconsin, 
Vanderbilt University, and Yale University.

Thanks to Cao Xiao, Bao Min, Zhou Yuren, Mo Hao and Li Songlin for their suggestions and discussions. 

\section*{Data Availability}

The data underlying this article will be shared on reasonable request to the corresponding author.

\bibliographystyle{mnras}
\bibliography{bicone_galaxy}

\begin{thebibliography}{}
\makeatletter
\relax
\def\mn@urlcharsother{\let\do\@makeother \do\$\do\&\do\#\do\^\do\_\do\%\do\~}
\def\mn@doi{\begingroup\mn@urlcharsother \@ifnextchar [ {\mn@doi@} {\mn@doi@[]}}
\def\mn@doi@[#1]#2{\def\@tempa{#1}\ifx\@tempa\@empty \href {http://dx.doi.org/#2} {doi:#2}\else \href {http://dx.doi.org/#2} {#1}\fi \endgroup}
\def\mn@eprint#1#2{\mn@eprint@#1:#2::\@nil}
\def\mn@eprint@arXiv#1{\href {http://arxiv.org/abs/#1} {{\tt arXiv:#1}}}
\def\mn@eprint@dblp#1{\href {http://dblp.uni-trier.de/rec/bibtex/#1.xml} {dblp:#1}}
\def\mn@eprint@#1:#2:#3:#4\@nil{\def\@tempa {#1}\def\@tempb {#2}\def\@tempc {#3}\ifx \@tempc \@empty \let \@tempc \@tempb \let \@tempb \@tempa \fi \ifx \@tempb \@empty \def\@tempb {arXiv}\fi \@ifundefined {mn@eprint@\@tempb}{\@tempb:\@tempc}{\expandafter \expandafter \csname mn@eprint@\@tempb\endcsname \expandafter{\@tempc}}}

\bibitem[\protect\citeauthoryear{{Abazajian} et~al.,}{{Abazajian} et~al.}{2004}]{2004AJ....128..502A}
{Abazajian} K.,  et~al., 2004, \mn@doi [\aj] {10.1086/421365}, \href {https://ui.adsabs.harvard.edu/abs/2004AJ....128..502A} {128, 502}

\bibitem[\protect\citeauthoryear{{Abdurro'uf} et~al.,}{{Abdurro'uf} et~al.}{2022}]{2022ApJS..259...35A}
{Abdurro'uf} et~al., 2022, \mn@doi [\apjs] {10.3847/1538-4365/ac4414}, \href {https://ui.adsabs.harvard.edu/abs/2022ApJS..259...35A} {259, 35}

\bibitem[\protect\citeauthoryear{{Abramowicz} \& {Fragile}}{{Abramowicz} \& {Fragile}}{2013}]{2013LRR....16....1A}
{Abramowicz} M.~A.,  {Fragile} P.~C.,  2013, \mn@doi [Living Reviews in Relativity] {10.12942/lrr-2013-1}, \href {https://ui.adsabs.harvard.edu/abs/2013LRR....16....1A} {16, 1}

\bibitem[\protect\citeauthoryear{{Alb{\'a}n} \& {Wylezalek}}{{Alb{\'a}n} \& {Wylezalek}}{2023}]{2023A&A...674A..85A}
{Alb{\'a}n} M.,  {Wylezalek} D.,  2023, \mn@doi [\aap] {10.1051/0004-6361/202245437}, \href {https://ui.adsabs.harvard.edu/abs/2023A&A...674A..85A} {674, A85}

\bibitem[\protect\citeauthoryear{{Aravindan}, {Liu}  \& {Canalizo}}{{Aravindan} et~al.}{2023}]{2023AAS...24136109A}
{Aravindan} A.,  {Liu} W.,   {Canalizo} G.,  2023, in American Astronomical Society Meeting Abstracts. p. 361.09

\bibitem[\protect\citeauthoryear{{Baldwin}, {Phillips}  \& {Terlevich}}{{Baldwin} et~al.}{1981}]{1981PASP...93....5B}
{Baldwin} J.~A.,  {Phillips} M.~M.,   {Terlevich} R.,  1981, \mn@doi [\pasp] {10.1086/130766}, \href {https://ui.adsabs.harvard.edu/abs/1981PASP...93....5B} {93, 5}

\bibitem[\protect\citeauthoryear{{Bao}, {Chen}, {Yuan}, {Shi}, {Bizyaev}, {Yu}, {Gu}  \& {Yu}}{{Bao} et~al.}{2019}]{2019MNRAS.490.3830B}
{Bao} M.,  {Chen} Y.-m.,  {Yuan} Q.-r.,  {Shi} Y.,  {Bizyaev} D.,  {Yu} X.-l.,  {Gu} Q.-s.,   {Yu} Y.,  2019, \mn@doi [\mnras] {10.1093/mnras/stz2892}, \href {https://ui.adsabs.harvard.edu/abs/2019MNRAS.490.3830B} {490, 3830}

\bibitem[\protect\citeauthoryear{{Bao} et~al.,}{{Bao} et~al.}{2021}]{2021MNRAS.505..191B}
{Bao} M.,  et~al., 2021, \mn@doi [\mnras] {10.1093/mnras/stab1201}, \href {https://ui.adsabs.harvard.edu/abs/2021MNRAS.505..191B} {505, 191}

\bibitem[\protect\citeauthoryear{{B{\"a}r}, {Weigel}, {Sartori}, {Oh}, {Koss}  \& {Schawinski}}{{B{\"a}r} et~al.}{2017}]{2017MNRAS.466.2879B}
{B{\"a}r} R.~E.,  {Weigel} A.~K.,  {Sartori} L.~F.,  {Oh} K.,  {Koss} M.,   {Schawinski} K.,  2017, \mn@doi [\mnras] {10.1093/mnras/stw3283}, \href {https://ui.adsabs.harvard.edu/abs/2017MNRAS.466.2879B} {466, 2879}

\bibitem[\protect\citeauthoryear{{Belfiore} et~al.,}{{Belfiore} et~al.}{2016}]{2016MNRAS.461.3111B}
{Belfiore} F.,  et~al., 2016, \mn@doi [\mnras] {10.1093/mnras/stw1234}, \href {https://ui.adsabs.harvard.edu/abs/2016MNRAS.461.3111B} {461, 3111}

\bibitem[\protect\citeauthoryear{{Belfiore} et~al.,}{{Belfiore} et~al.}{2019}]{2019AJ....158..160B}
{Belfiore} F.,  et~al., 2019, \mn@doi [\aj] {10.3847/1538-3881/ab3e4e}, \href {https://ui.adsabs.harvard.edu/abs/2019AJ....158..160B} {158, 160}

\bibitem[\protect\citeauthoryear{{Best} \& {Heckman}}{{Best} \& {Heckman}}{2012}]{2012MNRAS.421.1569B}
{Best} P.~N.,  {Heckman} T.~M.,  2012, \mn@doi [\mnras] {10.1111/j.1365-2966.2012.20414.x}, \href {https://ui.adsabs.harvard.edu/abs/2012MNRAS.421.1569B} {421, 1569}

\bibitem[\protect\citeauthoryear{{Bizyaev}, {Chen}, {Shi}, {Riffel}, {Riffel}, {Diamond-Stanic}  \& {Roy}}{{Bizyaev} et~al.}{2019}]{2019ApJ...882..145B}
{Bizyaev} D.,  {Chen} Y.-M.,  {Shi} Y.,  {Riffel} R.~A.,  {Riffel} R.,  {Diamond-Stanic} A.~M.,   {Roy} N.,  2019, \mn@doi [\apj] {10.3847/1538-4357/ab3406}, \href {https://ui.adsabs.harvard.edu/abs/2019ApJ...882..145B} {882, 145}

\bibitem[\protect\citeauthoryear{{Bizyaev}, {Chen}, {Shi}, {Roy}, {Riffel}, {Riffel}  \& {Fern{\'a}ndez-Trincado}}{{Bizyaev} et~al.}{2022}]{2022MNRAS.516.3092B}
{Bizyaev} D.,  {Chen} Y.-M.,  {Shi} Y.,  {Roy} N.,  {Riffel} R.,  {Riffel} R.~A.,   {Fern{\'a}ndez-Trincado} J.~G.,  2022, \mn@doi [\mnras] {10.1093/mnras/stac2439}, \href {https://ui.adsabs.harvard.edu/abs/2022MNRAS.516.3092B} {516, 3092}

\bibitem[\protect\citeauthoryear{{Blanton} \& {Roweis}}{{Blanton} \& {Roweis}}{2007}]{2007AJ....133..734B}
{Blanton} M.~R.,  {Roweis} S.,  2007, \mn@doi [\aj] {10.1086/510127}, \href {https://ui.adsabs.harvard.edu/abs/2007AJ....133..734B} {133, 734}

\bibitem[\protect\citeauthoryear{Blanton, Kazin, Muna, Weaver  \& Price-Whelan}{Blanton et~al.}{2011}]{Blanton_2011}
Blanton M.~R.,  Kazin E.,  Muna D.,  Weaver B.~A.,   Price-Whelan A.,  2011, \mn@doi [The Astronomical Journal] {10.1088/0004-6256/142/1/31}, 142, 31

\bibitem[\protect\citeauthoryear{{Blanton} et~al.,}{{Blanton} et~al.}{2017}]{2017AJ....154...28B}
{Blanton} M.~R.,  et~al., 2017, \mn@doi [\aj] {10.3847/1538-3881/aa7567}, \href {https://ui.adsabs.harvard.edu/abs/2017AJ....154...28B} {154, 28}

\bibitem[\protect\citeauthoryear{{Bolatto} et~al.,}{{Bolatto} et~al.}{2013}]{2013Natur.499..450B}
{Bolatto} A.~D.,  et~al., 2013, \mn@doi [\nat] {10.1038/nature12351}, \href {https://ui.adsabs.harvard.edu/abs/2013Natur.499..450B} {499, 450}

\bibitem[\protect\citeauthoryear{{Bower}, {Benson}, {Malbon}, {Helly}, {Frenk}, {Baugh}, {Cole}  \& {Lacey}}{{Bower} et~al.}{2006}]{2006MNRAS.370..645B}
{Bower} R.~G.,  {Benson} A.~J.,  {Malbon} R.,  {Helly} J.~C.,  {Frenk} C.~S.,  {Baugh} C.~M.,  {Cole} S.,   {Lacey} C.~G.,  2006, \mn@doi [\mnras] {10.1111/j.1365-2966.2006.10519.x}, \href {https://ui.adsabs.harvard.edu/abs/2006MNRAS.370..645B} {370, 645}

\bibitem[\protect\citeauthoryear{{Bruzual} \& {Charlot}}{{Bruzual} \& {Charlot}}{2003}]{2003MNRAS.344.1000B}
{Bruzual} G.,  {Charlot} S.,  2003, \mn@doi [\mnras] {10.1046/j.1365-8711.2003.06897.x}, \href {https://ui.adsabs.harvard.edu/abs/2003MNRAS.344.1000B} {344, 1000}

\bibitem[\protect\citeauthoryear{{Bundy} et~al.,}{{Bundy} et~al.}{2015}]{2015ApJ...798....7B}
{Bundy} K.,  et~al., 2015, \mn@doi [\apj] {10.1088/0004-637X/798/1/7}, \href {https://ui.adsabs.harvard.edu/abs/2015ApJ...798....7B} {798, 7}

\bibitem[\protect\citeauthoryear{{Butler} et~al.,}{{Butler} et~al.}{2021}]{2021ApJ...919....5B}
{Butler} K.~M.,  et~al., 2021, \mn@doi [\apj] {10.3847/1538-4357/ac0c7a}, \href {https://ui.adsabs.harvard.edu/abs/2021ApJ...919....5B} {919, 5}

\bibitem[\protect\citeauthoryear{Calzetti, Armus, Bohlin, Kinney, Koornneef  \& Storchi-Bergmann}{Calzetti et~al.}{2000}]{Calzetti_2000}
Calzetti D.,  Armus L.,  Bohlin R.~C.,  Kinney A.~L.,  Koornneef J.,   Storchi-Bergmann T.,  2000, \mn@doi [The Astrophysical Journal] {10.1086/308692}, 533, 682

\bibitem[\protect\citeauthoryear{{Cappellari} \& {Emsellem}}{{Cappellari} \& {Emsellem}}{2004}]{2004PASP..116..138C}
{Cappellari} M.,  {Emsellem} E.,  2004, \mn@doi [\pasp] {10.1086/381875}, \href {https://ui.adsabs.harvard.edu/abs/2004PASP..116..138C} {116, 138}

\bibitem[\protect\citeauthoryear{{Chabrier}}{{Chabrier}}{2003}]{2003PASP..115..763C}
{Chabrier} G.,  2003, \mn@doi [\pasp] {10.1086/376392}, \href {https://ui.adsabs.harvard.edu/abs/2003PASP..115..763C} {115, 763}

\bibitem[\protect\citeauthoryear{Chen, Tremonti, Heckman, Kauffmann, Weiner, Brinchmann  \& Wang}{Chen et~al.}{2010}]{Chen_2010}
Chen Y.-M.,  Tremonti C.~A.,  Heckman T.~M.,  Kauffmann G.,  Weiner B.~J.,  Brinchmann J.,   Wang J.,  2010, \mn@doi [The Astronomical Journal] {10.1088/0004-6256/140/2/445}, 140, 445

\bibitem[\protect\citeauthoryear{{Cheung} et~al.,}{{Cheung} et~al.}{2016}]{2016Natur.533..504C}
{Cheung} E.,  et~al., 2016, \mn@doi [\nat] {10.1038/nature18006}, \href {https://ui.adsabs.harvard.edu/abs/2016Natur.533..504C} {533, 504}

\bibitem[\protect\citeauthoryear{{Chevalier} \& {Clegg}}{{Chevalier} \& {Clegg}}{1985}]{1985Natur.317...44C}
{Chevalier} R.~A.,  {Clegg} A.~W.,  1985, \mn@doi [\nat] {10.1038/317044a0}, \href {https://ui.adsabs.harvard.edu/abs/1985Natur.317...44C} {317, 44}

\bibitem[\protect\citeauthoryear{{Cid Fernandes}, {Stasi{\'n}ska}, {Schlickmann}, {Mateus}, {Vale Asari}, {Schoenell}  \& {Sodr{\'e}}}{{Cid Fernandes} et~al.}{2010}]{2010MNRAS.403.1036C}
{Cid Fernandes} R.,  {Stasi{\'n}ska} G.,  {Schlickmann} M.~S.,  {Mateus} A.,  {Vale Asari} N.,  {Schoenell} W.,   {Sodr{\'e}} L.,  2010, \mn@doi [\mnras] {10.1111/j.1365-2966.2009.16185.x}, \href {https://ui.adsabs.harvard.edu/abs/2010MNRAS.403.1036C} {403, 1036}

\bibitem[\protect\citeauthoryear{{Comerford} et~al.,}{{Comerford} et~al.}{2020}]{2020ApJ...901..159C}
{Comerford} J.~M.,  et~al., 2020, \mn@doi [\apj] {10.3847/1538-4357/abb2ae}, \href {https://ui.adsabs.harvard.edu/abs/2020ApJ...901..159C} {901, 159}

\bibitem[\protect\citeauthoryear{{Crenshaw}, {Fischer}, {Kraemer}  \& {Schmitt}}{{Crenshaw} et~al.}{2015}]{2015ApJ...799...83C}
{Crenshaw} D.~M.,  {Fischer} T.~C.,  {Kraemer} S.~B.,   {Schmitt} H.~R.,  2015, \mn@doi [\apj] {10.1088/0004-637X/799/1/83}, \href {https://ui.adsabs.harvard.edu/abs/2015ApJ...799...83C} {799, 83}

\bibitem[\protect\citeauthoryear{{Croom} et~al.,}{{Croom} et~al.}{2012}]{2012MNRAS.421..872C}
{Croom} S.~M.,  et~al., 2012, \mn@doi [\mnras] {10.1111/j.1365-2966.2011.20365.x}, \href {https://ui.adsabs.harvard.edu/abs/2012MNRAS.421..872C} {421, 872}

\bibitem[\protect\citeauthoryear{{Croton} et~al.,}{{Croton} et~al.}{2006}]{2006MNRAS.365...11C}
{Croton} D.~J.,  et~al., 2006, \mn@doi [\mnras] {10.1111/j.1365-2966.2005.09675.x}, \href {https://ui.adsabs.harvard.edu/abs/2006MNRAS.365...11C} {365, 11}

\bibitem[\protect\citeauthoryear{{Debuhr}, {Quataert}  \& {Ma}}{{Debuhr} et~al.}{2012}]{2012MNRAS.420.2221D}
{Debuhr} J.,  {Quataert} E.,   {Ma} C.-P.,  2012, \mn@doi [\mnras] {10.1111/j.1365-2966.2011.20187.x}, \href {https://ui.adsabs.harvard.edu/abs/2012MNRAS.420.2221D} {420, 2221}

\bibitem[\protect\citeauthoryear{{Dirks}, {Dettmar}, {Bomans}, {Kamphuis}  \& {Schilling}}{{Dirks} et~al.}{2023}]{2023A&A...678A..84D}
{Dirks} L.,  {Dettmar} R.~J.,  {Bomans} D.~J.,  {Kamphuis} P.,   {Schilling} U.,  2023, \mn@doi [\aap] {10.1051/0004-6361/202245679}, \href {https://ui.adsabs.harvard.edu/abs/2023A&A...678A..84D} {678, A84}

\bibitem[\protect\citeauthoryear{{Engelbracht} et~al.,}{{Engelbracht} et~al.}{2006}]{2006ApJ...642L.127E}
{Engelbracht} C.~W.,  et~al., 2006, \mn@doi [\apjl] {10.1086/504590}, \href {https://ui.adsabs.harvard.edu/abs/2006ApJ...642L.127E} {642, L127}

\bibitem[\protect\citeauthoryear{{Faucher-Gigu{\`e}re} \& {Quataert}}{{Faucher-Gigu{\`e}re} \& {Quataert}}{2012}]{2012MNRAS.425..605F}
{Faucher-Gigu{\`e}re} C.-A.,  {Quataert} E.,  2012, \mn@doi [\mnras] {10.1111/j.1365-2966.2012.21512.x}, \href {https://ui.adsabs.harvard.edu/abs/2012MNRAS.425..605F} {425, 605}

\bibitem[\protect\citeauthoryear{{Fukugita}, {Ichikawa}, {Gunn}, {Doi}, {Shimasaku}  \& {Schneider}}{{Fukugita} et~al.}{1996}]{1996AJ....111.1748F}
{Fukugita} M.,  {Ichikawa} T.,  {Gunn} J.~E.,  {Doi} M.,  {Shimasaku} K.,   {Schneider} D.~P.,  1996, \mn@doi [\aj] {10.1086/117915}, \href {https://ui.adsabs.harvard.edu/abs/1996AJ....111.1748F} {111, 1748}

\bibitem[\protect\citeauthoryear{{Gebhardt} et~al.,}{{Gebhardt} et~al.}{2000}]{2000ApJ...539L..13G}
{Gebhardt} K.,  et~al., 2000, \mn@doi [\apjl] {10.1086/312840}, \href {https://ui.adsabs.harvard.edu/abs/2000ApJ...539L..13G} {539, L13}

\bibitem[\protect\citeauthoryear{{Gunn} et~al.,}{{Gunn} et~al.}{1998}]{1998AJ....116.3040G}
{Gunn} J.~E.,  et~al., 1998, \mn@doi [\aj] {10.1086/300645}, \href {https://ui.adsabs.harvard.edu/abs/1998AJ....116.3040G} {116, 3040}

\bibitem[\protect\citeauthoryear{{Gunn} et~al.,}{{Gunn} et~al.}{2006}]{2006AJ....131.2332G}
{Gunn} J.~E.,  et~al., 2006, \mn@doi [\aj] {10.1086/500975}, \href {https://ui.adsabs.harvard.edu/abs/2006AJ....131.2332G} {131, 2332}

\bibitem[\protect\citeauthoryear{{Heckman}}{{Heckman}}{1980}]{1980A&A....87..152H}
{Heckman} T.~M.,  1980, \aap, \href {https://ui.adsabs.harvard.edu/abs/1980A&A....87..152H} {87, 152}

\bibitem[\protect\citeauthoryear{{Heckman}}{{Heckman}}{2002}]{2002ASPC..254..292H}
{Heckman} T.~M.,  2002, in {Mulchaey} J.~S.,  {Stocke} J.~T.,  eds,  Astronomical Society of the Pacific Conference Series Vol. 254, Extragalactic Gas at Low Redshift. p.~292 (\mn@eprint {arXiv} {astro-ph/0107438}), \mn@doi{10.48550/arXiv.astro-ph/0107438}

\bibitem[\protect\citeauthoryear{{Heckman}}{{Heckman}}{2003}]{2003RMxAC..17...47H}
{Heckman} T.~M.,  2003, in {Avila-Reese} V.,  {Firmani} C.,  {Frenk} C.~S.,   {Allen} C.,  eds,  Revista Mexicana de Astronomia y Astrofisica Conference Series Vol. 17, Revista Mexicana de Astronomia y Astrofisica Conference Series. pp 47--55

\bibitem[\protect\citeauthoryear{{Heckman}, {Armus}  \& {Miley}}{{Heckman} et~al.}{1990}]{1990ApJS...74..833H}
{Heckman} T.~M.,  {Armus} L.,   {Miley} G.~K.,  1990, \mn@doi [\apjs] {10.1086/191522}, \href {https://ui.adsabs.harvard.edu/abs/1990ApJS...74..833H} {74, 833}

\bibitem[\protect\citeauthoryear{{Hopkins}, {Hernquist}, {Hayward}  \& {Narayanan}}{{Hopkins} et~al.}{2012}]{2012MNRAS.425.1121H}
{Hopkins} P.~F.,  {Hernquist} L.,  {Hayward} C.~C.,   {Narayanan} D.,  2012, \mn@doi [\mnras] {10.1111/j.1365-2966.2012.21449.x}, \href {https://ui.adsabs.harvard.edu/abs/2012MNRAS.425.1121H} {425, 1121}

\bibitem[\protect\citeauthoryear{{Juneau} et~al.,}{{Juneau} et~al.}{2022}]{2022ApJ...925..203J}
{Juneau} S.,  et~al., 2022, \mn@doi [\apj] {10.3847/1538-4357/ac425f}, \href {https://ui.adsabs.harvard.edu/abs/2022ApJ...925..203J} {925, 203}

\bibitem[\protect\citeauthoryear{{Kauffmann} et~al.,}{{Kauffmann} et~al.}{2003}]{2003MNRAS.346.1055K}
{Kauffmann} G.,  et~al., 2003, \mn@doi [\mnras] {10.1111/j.1365-2966.2003.07154.x}, \href {https://ui.adsabs.harvard.edu/abs/2003MNRAS.346.1055K} {346, 1055}

\bibitem[\protect\citeauthoryear{{Keel}}{{Keel}}{1983}]{1983ApJS...52..229K}
{Keel} W.~C.,  1983, \mn@doi [\apjs] {10.1086/190866}, \href {https://ui.adsabs.harvard.edu/abs/1983ApJS...52..229K} {52, 229}

\bibitem[\protect\citeauthoryear{Kennicutt}{Kennicutt}{1998}]{doi:10.1146/annurev.astro.36.1.189}
Kennicutt R.~C.,  1998, \mn@doi [Annual Review of Astronomy and Astrophysics] {10.1146/annurev.astro.36.1.189}, 36, 189

\bibitem[\protect\citeauthoryear{Kewley, Dopita, Sutherland, Heisler  \& Trevena}{Kewley et~al.}{2001}]{Kewley_2001}
Kewley L.~J.,  Dopita M.~A.,  Sutherland R.~S.,  Heisler C.~A.,   Trevena J.,  2001, \mn@doi [The Astrophysical Journal] {10.1086/321545}, 556, 121

\bibitem[\protect\citeauthoryear{King \& Pounds}{King \& Pounds}{2015}]{doi:10.1146/annurev-astro-082214-122316}
King A.,  Pounds K.,  2015, \mn@doi [Annual Review of Astronomy and Astrophysics] {10.1146/annurev-astro-082214-122316}, 53, 115

\bibitem[\protect\citeauthoryear{Kolmogorov}{Kolmogorov}{1933}]{1572824501049496320}
Kolmogorov A.,  1933, Giorn Dell'inst Ital Degli Att, 4, 89

\bibitem[\protect\citeauthoryear{{Krajnovi{\'c}}, {Cappellari}, {de Zeeuw}  \& {Copin}}{{Krajnovi{\'c}} et~al.}{2006}]{2006MNRAS.366..787K}
{Krajnovi{\'c}} D.,  {Cappellari} M.,  {de Zeeuw} P.~T.,   {Copin} Y.,  2006, \mn@doi [\mnras] {10.1111/j.1365-2966.2005.09902.x}, \href {https://ui.adsabs.harvard.edu/abs/2006MNRAS.366..787K} {366, 787}

\bibitem[\protect\citeauthoryear{{LaMassa}, {Heckman}, {Ptak}, {Hornschemeier}, {Martins}, {Sonnentrucker}  \& {Tremonti}}{{LaMassa} et~al.}{2009}]{2009ApJ...705..568L}
{LaMassa} S.~M.,  {Heckman} T.~M.,  {Ptak} A.,  {Hornschemeier} A.,  {Martins} L.,  {Sonnentrucker} P.,   {Tremonti} C.,  2009, \mn@doi [\apj] {10.1088/0004-637X/705/1/568}, \href {https://ui.adsabs.harvard.edu/abs/2009ApJ...705..568L} {705, 568}

\bibitem[\protect\citeauthoryear{{Law} et~al.,}{{Law} et~al.}{2015}]{2015AJ....150...19L}
{Law} D.~R.,  et~al., 2015, \mn@doi [\aj] {10.1088/0004-6256/150/1/19}, \href {https://ui.adsabs.harvard.edu/abs/2015AJ....150...19L} {150, 19}

\bibitem[\protect\citeauthoryear{{Lupton}, {Gunn}, {Ivezi{\'c}}, {Knapp}  \& {Kent}}{{Lupton} et~al.}{2001}]{2001ASPC..238..269L}
{Lupton} R.,  {Gunn} J.~E.,  {Ivezi{\'c}} Z.,  {Knapp} G.~R.,   {Kent} S.,  2001, in {Harnden} F.~R. J.,  {Primini} F.~A.,   {Payne} H.~E.,  eds,  Astronomical Society of the Pacific Conference Series Vol. 238, Astronomical Data Analysis Software and Systems X. p.~269 (\mn@eprint {arXiv} {astro-ph/0101420}), \mn@doi{10.48550/arXiv.astro-ph/0101420}

\bibitem[\protect\citeauthoryear{{Martig}, {Bournaud}, {Teyssier}  \& {Dekel}}{{Martig} et~al.}{2009}]{2009ApJ...707..250M}
{Martig} M.,  {Bournaud} F.,  {Teyssier} R.,   {Dekel} A.,  2009, \mn@doi [\apj] {10.1088/0004-637X/707/1/250}, \href {https://ui.adsabs.harvard.edu/abs/2009ApJ...707..250M} {707, 250}

\bibitem[\protect\citeauthoryear{{Mutchler} et~al.,}{{Mutchler} et~al.}{2007}]{2007PASP..119....1M}
{Mutchler} M.,  et~al., 2007, \mn@doi [\pasp] {10.1086/511160}, \href {https://ui.adsabs.harvard.edu/abs/2007PASP..119....1M} {119, 1}

\bibitem[\protect\citeauthoryear{{Nagar} \& {Wilson}}{{Nagar} \& {Wilson}}{1999}]{1999ApJ...516...97N}
{Nagar} N.~M.,  {Wilson} A.~S.,  1999, \mn@doi [\apj] {10.1086/307109}, \href {https://ui.adsabs.harvard.edu/abs/1999ApJ...516...97N} {516, 97}

\bibitem[\protect\citeauthoryear{{Nakai}, {Hayashi}, {Handa}, {Sofue}, {Hasegawa}  \& {Sasaki}}{{Nakai} et~al.}{1987}]{1987PASJ...39..685N}
{Nakai} N.,  {Hayashi} M.,  {Handa} T.,  {Sofue} Y.,  {Hasegawa} T.,   {Sasaki} M.,  1987, \pasj, \href {https://ui.adsabs.harvard.edu/abs/1987PASJ...39..685N} {39, 685}

\bibitem[\protect\citeauthoryear{{Padilla} \& {Strauss}}{{Padilla} \& {Strauss}}{2008}]{2008MNRAS.388.1321P}
{Padilla} N.~D.,  {Strauss} M.~A.,  2008, \mn@doi [\mnras] {10.1111/j.1365-2966.2008.13480.x}, \href {https://ui.adsabs.harvard.edu/abs/2008MNRAS.388.1321P} {388, 1321}

\bibitem[\protect\citeauthoryear{{Pounds} \& {Reeves}}{{Pounds} \& {Reeves}}{2009}]{2009MNRAS.397..249P}
{Pounds} K.~A.,  {Reeves} J.~N.,  2009, \mn@doi [\mnras] {10.1111/j.1365-2966.2009.14971.x}, \href {https://ui.adsabs.harvard.edu/abs/2009MNRAS.397..249P} {397, 249}

\bibitem[\protect\citeauthoryear{{Rautio}, {Watkins}, {Comer{\'o}n}, {Salo}, {D{\'\i}az-Garc{\'\i}a}  \& {Janz}}{{Rautio} et~al.}{2022}]{2022A&A...659A.153R}
{Rautio} R.~P.~V.,  {Watkins} A.~E.,  {Comer{\'o}n} S.,  {Salo} H.,  {D{\'\i}az-Garc{\'\i}a} S.,   {Janz} J.,  2022, \mn@doi [\aap] {10.1051/0004-6361/202142440}, \href {https://ui.adsabs.harvard.edu/abs/2022A&A...659A.153R} {659, A153}

\bibitem[\protect\citeauthoryear{{Roy} et~al.,}{{Roy} et~al.}{2018}]{2018ApJ...869..117R}
{Roy} N.,  et~al., 2018, \mn@doi [\apj] {10.3847/1538-4357/aaee72}, \href {https://ui.adsabs.harvard.edu/abs/2018ApJ...869..117R} {869, 117}

\bibitem[\protect\citeauthoryear{{Roy} et~al.,}{{Roy} et~al.}{2021a}]{2021ApJ...913...33R}
{Roy} N.,  et~al., 2021a, \mn@doi [\apj] {10.3847/1538-4357/abf1e6}, \href {https://ui.adsabs.harvard.edu/abs/2021ApJ...913...33R} {913, 33}

\bibitem[\protect\citeauthoryear{{Roy} et~al.,}{{Roy} et~al.}{2021b}]{2021ApJ...922..230R}
{Roy} N.,  et~al., 2021b, \mn@doi [\apj] {10.3847/1538-4357/ac24a0}, \href {https://ui.adsabs.harvard.edu/abs/2021ApJ...922..230R} {922, 230}

\bibitem[\protect\citeauthoryear{{Russell} et~al.,}{{Russell} et~al.}{2019}]{2019MNRAS.490.3025R}
{Russell} H.~R.,  et~al., 2019, \mn@doi [\mnras] {10.1093/mnras/stz2719}, \href {https://ui.adsabs.harvard.edu/abs/2019MNRAS.490.3025R} {490, 3025}

\bibitem[\protect\citeauthoryear{{Sakamoto}, {Aalto}, {Combes}, {Evans}  \& {Peck}}{{Sakamoto} et~al.}{2014}]{2014ApJ...797...90S}
{Sakamoto} K.,  {Aalto} S.,  {Combes} F.,  {Evans} A.,   {Peck} A.,  2014, \mn@doi [\apj] {10.1088/0004-637X/797/2/90}, \href {https://ui.adsabs.harvard.edu/abs/2014ApJ...797...90S} {797, 90}

\bibitem[\protect\citeauthoryear{{S{\'a}nchez-Bl{\'a}zquez} et~al.,}{{S{\'a}nchez-Bl{\'a}zquez} et~al.}{2006}]{2006MNRAS.371..703S}
{S{\'a}nchez-Bl{\'a}zquez} P.,  et~al., 2006, \mn@doi [\mnras] {10.1111/j.1365-2966.2006.10699.x}, \href {https://ui.adsabs.harvard.edu/abs/2006MNRAS.371..703S} {371, 703}

\bibitem[\protect\citeauthoryear{{S{\'a}nchez} et~al.,}{{S{\'a}nchez} et~al.}{2012}]{2012A&A...538A...8S}
{S{\'a}nchez} S.~F.,  et~al., 2012, \mn@doi [\aap] {10.1051/0004-6361/201117353}, \href {https://ui.adsabs.harvard.edu/abs/2012A&A...538A...8S} {538, A8}

\bibitem[\protect\citeauthoryear{{S{\'a}nchez} et~al.,}{{S{\'a}nchez} et~al.}{2016}]{2016RMxAA..52..171S}
{S{\'a}nchez} S.~F.,  et~al., 2016, \mn@doi [\rmxaa] {10.48550/arXiv.1602.01830}, \href {https://ui.adsabs.harvard.edu/abs/2016RMxAA..52..171S} {52, 171}

\bibitem[\protect\citeauthoryear{{Sarzi} et~al.,}{{Sarzi} et~al.}{2006}]{2006MNRAS.366.1151S}
{Sarzi} M.,  et~al., 2006, \mn@doi [\mnras] {10.1111/j.1365-2966.2005.09839.x}, \href {https://ui.adsabs.harvard.edu/abs/2006MNRAS.366.1151S} {366, 1151}

\bibitem[\protect\citeauthoryear{{Sarzi} et~al.,}{{Sarzi} et~al.}{2010}]{2010MNRAS.402.2187S}
{Sarzi} M.,  et~al., 2010, \mn@doi [\mnras] {10.1111/j.1365-2966.2009.16039.x}, \href {https://ui.adsabs.harvard.edu/abs/2010MNRAS.402.2187S} {402, 2187}

\bibitem[\protect\citeauthoryear{Schmitt \& Kinney}{Schmitt \& Kinney}{2002}]{SCHMITT2002231}
Schmitt H.,  Kinney A.,  2002, \mn@doi [New Astronomy Reviews] {https://doi.org/10.1016/S1387-6473(01)00186-5}, 46, 231

\bibitem[\protect\citeauthoryear{{Silk}}{{Silk}}{1997}]{1997ApJ...481..703S}
{Silk} J.,  1997, \mn@doi [\apj] {10.1086/304073}, \href {https://ui.adsabs.harvard.edu/abs/1997ApJ...481..703S} {481, 703}

\bibitem[\protect\citeauthoryear{{Smee} et~al.,}{{Smee} et~al.}{2013}]{2013AJ....146...32S}
{Smee} S.~A.,  et~al., 2013, \mn@doi [\aj] {10.1088/0004-6256/146/2/32}, \href {https://ui.adsabs.harvard.edu/abs/2013AJ....146...32S} {146, 32}

\bibitem[\protect\citeauthoryear{Smirnov}{Smirnov}{1948}]{1363670321263386240}
Smirnov N.,  1948, \mn@doi [The Annals of Mathematical Statistics] {10.1214/aoms/1177730256}, 19, 279

\bibitem[\protect\citeauthoryear{{Smith} et~al.,}{{Smith} et~al.}{2002}]{2002AJ....123.2121S}
{Smith} J.~A.,  et~al., 2002, \mn@doi [\aj] {10.1086/339311}, \href {https://ui.adsabs.harvard.edu/abs/2002AJ....123.2121S} {123, 2121}

\bibitem[\protect\citeauthoryear{{Springel} \& {Hernquist}}{{Springel} \& {Hernquist}}{2003}]{2003MNRAS.339..289S}
{Springel} V.,  {Hernquist} L.,  2003, \mn@doi [\mnras] {10.1046/j.1365-8711.2003.06206.x}, \href {https://ui.adsabs.harvard.edu/abs/2003MNRAS.339..289S} {339, 289}

\bibitem[\protect\citeauthoryear{{Springel}, {Di Matteo}  \& {Hernquist}}{{Springel} et~al.}{2005}]{2005MNRAS.361..776S}
{Springel} V.,  {Di Matteo} T.,   {Hernquist} L.,  2005, \mn@doi [\mnras] {10.1111/j.1365-2966.2005.09238.x}, \href {https://ui.adsabs.harvard.edu/abs/2005MNRAS.361..776S} {361, 776}

\bibitem[\protect\citeauthoryear{{Strickland} \& {Heckman}}{{Strickland} \& {Heckman}}{2007}]{2007ApJ...658..258S}
{Strickland} D.~K.,  {Heckman} T.~M.,  2007, \mn@doi [\apj] {10.1086/511174}, \href {https://ui.adsabs.harvard.edu/abs/2007ApJ...658..258S} {658, 258}

\bibitem[\protect\citeauthoryear{{Tanner} \& {Weaver}}{{Tanner} \& {Weaver}}{2022}]{Tanner2022AGNoutflow}
{Tanner} R.,  {Weaver} K.~A.,  2022, \mn@doi [\aj] {10.3847/1538-3881/ac4d23}, \href {https://ui.adsabs.harvard.edu/abs/2022AJ....163..134T} {163, 134}

\bibitem[\protect\citeauthoryear{{Tumlinson}, {Peeples}  \& {Werk}}{{Tumlinson} et~al.}{2017}]{2017ARA&A..55..389T}
{Tumlinson} J.,  {Peeples} M.~S.,   {Werk} J.~K.,  2017, \mn@doi [\araa] {10.1146/annurev-astro-091916-055240}, \href {https://ui.adsabs.harvard.edu/abs/2017ARA&A..55..389T} {55, 389}

\bibitem[\protect\citeauthoryear{{Veilleux}, {Cecil}  \& {Bland-Hawthorn}}{{Veilleux} et~al.}{2005}]{2005ARA&A..43..769V}
{Veilleux} S.,  {Cecil} G.,   {Bland-Hawthorn} J.,  2005, \mn@doi [\araa] {10.1146/annurev.astro.43.072103.150610}, \href {https://ui.adsabs.harvard.edu/abs/2005ARA&A..43..769V} {43, 769}

\bibitem[\protect\citeauthoryear{{Wake} et~al.,}{{Wake} et~al.}{2017}]{2017AJ....154...86W}
{Wake} D.~A.,  et~al., 2017, \mn@doi [\aj] {10.3847/1538-3881/aa7ecc}, \href {https://ui.adsabs.harvard.edu/abs/2017AJ....154...86W} {154, 86}

\bibitem[\protect\citeauthoryear{{Weiner} et~al.,}{{Weiner} et~al.}{2009}]{2009ApJ...692..187W}
{Weiner} B.~J.,  et~al., 2009, \mn@doi [\apj] {10.1088/0004-637X/692/1/187}, \href {https://ui.adsabs.harvard.edu/abs/2009ApJ...692..187W} {692, 187}

\bibitem[\protect\citeauthoryear{{Yan} et~al.,}{{Yan} et~al.}{2016}]{2016AJ....151....8Y}
{Yan} R.,  et~al., 2016, \mn@doi [\aj] {10.3847/0004-6256/151/1/8}, \href {https://ui.adsabs.harvard.edu/abs/2016AJ....151....8Y} {151, 8}

\makeatother
\end{thebibliography}

\label{lastpage}
\end{document}